\documentclass[11pt,a4paper]{article}

\pdfoutput=1
\usepackage[left=2.5cm,right=2.5cm,top=2.5cm,bottom=2.5cm]{geometry}
\usepackage[onehalfspacing]{setspace}
    \usepackage[pdftex,     % sets up hyperref to use pdftex driver
            plainpages=false,   % allows page i and 1 to exist in the same document
            breaklinks=true,    % link texts can be broken at the end of line
            colorlinks=true
           ]{hyperref} 
\usepackage{thumbpdf}
\usepackage{amsmath}
\usepackage{amssymb}
\usepackage{array}
\usepackage{bbold}
\usepackage{cancel}
\usepackage[hang,small]{caption2}
\usepackage{subfigure}
\usepackage{verbatim}
\usepackage{slashed}
\usepackage{graphicx}
\usepackage{url}

\newcommand{\be}{\begin{equation}}
\newcommand{\ee}{\end{equation}}

\makeatletter
\def\maketag@@@#1{\hbox{\m@th\normalfont\normalsize#1}}
\makeatother

\begin{document}

\begin{titlepage}

\vspace*{1cm}

\begin{center}{
\bfseries\LARGE
WiLE: a Mathematica package for weak coupling expansion of Wilson loops in ABJ(M) theory}\\[8mm]
M.~Preti \footnote{E-mail: \texttt{michelangelo.preti@desy.de}}\\[1mm]
\end{center}

\vspace*{0.50cm}

\centerline{\itshape
DESY Hamburg, Theory Group, Notkestra\ss e 85, 22607 Hamburg, Germany}

\vspace*{1cm}
\begin{abstract}
We present \texttt{WiLE}, a \textit{Mathematica}\textsuperscript{\textregistered} package designed to perform the weak coupling expansion of any Wilson loop in ABJ(M) theory at arbitrary perturbative order. For a given set of fields on the loop and internal vertices, the package displays all the possible Feynman diagrams and their integral representations. The user can also choose to exclude non planar diagrams, tadpoles and self-energies. Through the use of interactive input windows, the package should be easily accessible to users with little or no previous experience. The package manual provides some pedagogical examples and the computation of all ladder diagrams at three-loop relevant for the cusp anomalous dimension in ABJ(M). The latter application gives also support to some recent results computed in different contexts.
\end{abstract}
\end{titlepage}

\section*{Program Summary}\noindent
\textbf{Author:} Michelangelo Preti\\\noindent
\textbf{Title:} WiLE\\\noindent
\textbf{Licence:} GNU LGPL\\\noindent
\textbf{Programming language / external routines:} Wolfram Mathematica version 8.0 or higher\\\noindent
\textbf{Operating system:} cross-platform\\\noindent
\textbf{RAM:} $>1$ GB RAM recommended\\\noindent
\textbf{Current version:} 1.0.0\\\noindent
\textbf{Nature of the problem:} In ABJ(M) theory one can define bosonic and fermionic Wilson loops. In order to study those operators using the standard perturbation theory techniques, one has to expand the Wilson loops at weak coupling and perform all the Wick contractions. Since the theory possesses a quiver structure and its Lagrangian contains many different vertices, this procedure could be very involved. Then we propose a Mathematica package to automatize this computation. \\\noindent
\textbf{Solution method:} The aim of the package \texttt{WiLE} is to obtain the weak coupling expansion of the Wilson loops in terms of Feynman diagrams. Given a set of initial data, as the set of vertices involved in the diagram and the fields contributing in the Wilson loop expansion, the package computes the related diagrams and their integral representations. \\\noindent
\textbf{Web page:}  \url{https://github.com/miciosca/WiLE}\\\noindent
\textbf{Contact:} \url{michelangelo.preti@desy.de} --- for bugs, possible improvements and questions

\section{Introduction}

Wilson loops are one of the most general gauge invariant observables in gauge theories and they play an important role in studying their general structure, both at perturbative and non-perturbative level. Depending on the contour, they encode different physical quantities, such as the potential between heavy moving probes or some aspects of scattering amplitudes of charged particles \cite{Polyakov:1980ca,Korchemsky:1991zp,Korchemsky:1993uz}. Furthermore, they are the building blocks of lattice formulation of any gauge theory, where they can be also used to study non-perturbative phenomenons. Remarkably, in the AdS/CFT framework \cite{Maldacena:1997re,Witten:1998qj,Gubser:1998bc}, a new interest for Wilson loop in  supersymmetric theories has been triggered. Indeed, they provide a rich class of BPS observables in supersymmetric gauge theories playing a central role in testing the correspondence itself, due to their relation with  the fundamental string or brane configurations present in the dual theory \cite{Maldacena:1998im,Rey:1998ik}. For example, whenever their vev is exactly computable (summing the perturbative series or through non perturbative techniques as supersymmetric localization or integrability), they provide non-trivial functions of the coupling which interpolate between weak and strong coupling regime. This allows for exact tests of the AdS/CFT correspondence \cite{Erickson:2000af,Drukker:2000rr}.

In $\mathcal{N}=4$ Super Yang-Mills theory (SYM), it was found a complete equivalence between scattering amplitudes in the planar limit and a particular class of null-polygonal Wilson loops in twistor space \cite{Mason:2010yk}. Other similar relation was proposed in \cite{ Drummond:2007cf,CaronHuot:2010ek}. Furthermore, dualities between Wilson loops and scattering amplitudes at weak and strong coupling have played a crucial role in understanding the integrable structure underlying both the gauge theory and its string dual (see \cite{Alday:2008yw,Schabinger:2011kb,Adamo:2011pv}). Correlation functions of Wilson loops and local operators was also computed exactly using supersymmetric localization and perturbation theory \cite{Giombi:2009ds,Giombi:2012ep,Bonini:2014vta} (see \cite{Preti:2016hhk} for a review). Recently, also Wilson loops with operator insertions were successfully studied both at weak and strong coupling using integrability \cite{Drukker:2012de,Correa:2012hh} and localization techniques \cite{Bonini:2015fng}, enlightening their role in the context of defect CFTs \cite{Cooke:2017qgm,Giombi:2017cqn}. Similar properties have also emerged in other superconformal theories as $\mathcal{N}=6$ Super Chern-Simons-matter (or ABJ(M)) theory in three dimensions \cite{Aharony:2008ug,Aharony:2008gk}.

Bosonic Wilson loops in ABJ(M) theory with gauge group $U(N)\times U(M)$ can be defined \cite{Drukker:2008zx} as the holonomy of a generalized gauge connection. This includes the coupling with scalars $C$ and $\bar C$ via a matrix  locally defined along the path. When the path is a maximal circle of $S^2$ or an infinite line, the Wilson loop is a 1/6 BPS operator (see also \cite{Drukker:2008zx,Chen:2008bp,Rey:2008bh}). Fermionic Wilson loops can be constructed turning on local couplings to the fermions $\psi$ and $\bar \psi$, generalizing the operator to the holonomy of a superconnection of the $U(N|M)$ supergroup \cite{Drukker:2009hy}. The supersymmetry of the latter is enhanced to 1/2 BPS on a  maximal circle or an infinite line contours (see also \cite{Lee:2010hk,Berenstein:2008dc}). 
The fermionic loop operator is cohomologically equivalent to a linear combination of bosonic Wilson loops \cite{Drukker:2009hy}. Perturbative computations for bosonic and fermionic Wilson loops have provided several results as the circular Wilson loop \cite{Bianchi:2013zda,Griguolo:2013sma,Bianchi:2013rma}, the cusp anomalous dimension \cite{Griguolo:2012iq,Bianchi:2014ada,Bonini:2016fnc} and the Bremsstrahlung function \cite{Bianchi:2014laa,Bianchi:2017svd,Bianchi:2017ozk}. These calculations are quite involved then, in order to simplify and automatize the weak coupling expansion in terms of Feynman diagrams, we have developed the \textit{Mathematica}\textsuperscript{\textregistered} package \texttt{WiLE}. 

The purpose of the package \texttt{WiLE} is to obtain the weak coupling expansion of the Wilson loop in terms of Feynman diagrams and their related integral representations. The package includes three functions: \texttt{WiLE}, \texttt{WiLEFullorder} and \texttt{WiLESimplify}. The first function, for a given set of input data as the number and the type of interaction vertices and fields on the loop\footnote{Expanding the Wilson loop operator up to a perturbative order $\ell$, we will end up with the product of $\ell$ (super-)connections. Each connection contributes as a gauge field, a fermion or a couple of scalars with their positions along the loop fixed by the path-ordering.}, draws the related Feynman diagrams and shows the integral representations of any diagram in terms of propagators, position vectors and local couplings of the fields.
The \texttt{WiLEFullorder} function gives the same output of the first one but it groups the complete expansion for a fixed perturbative order. Many other options can be chosen: the relevant diagrams in the large $N$ and $M$ limit can be selected, or particular classes as tadpoles or self-energies can be excluded. Finally, the \texttt{WiLESimplify} function provides the simplification of the output if needed.

The paper is organized as follows. In section 2 we review the bosonic and fermionic Wilson loops and their perturbative expansions. In section 3 we present in detail the manual of the \texttt{WiLE} package, while in section 4 we show an explicit applications. Few appendices follow, which contain conventions, some details of ABJ(M) action and Feynman rules and a detailed legend to interpret the package output.

\section{Wilson loops in ABJ(M) theory}\label{sec:susyWLABJM}

In the $AdS_4/CFT_3$ correspondence framework, the Wilson loops are dual to  
string surfaces \cite{Maldacena:1998im,Rey:1998ik}.
Since in the most symmetric case the string preserves 
half of the supercharges of the vacuum, the dual Wilson loop operator in the field theory 
must be 1/2 BPS. In analogy with the $\mathcal{N}=4$ SYM theory, the 
most natural choice is to consider a bosonic Wilson loop as the one defines in the $\mathcal{N}=2,3$ 
superconformal Chern-Simons-matter theories \cite{Gaiotto:2007qi}.
This loop was studied in the context of ABJ(M) in \cite{Drukker:2008zx,Chen:2008bp,Rey:2008bh} but, since it preserves 
only 1/6 of the supercharges (if it lie on a straight line or circular contour), it cannot be a viable candidate to be the dual of the 
classical string. In \cite{Drukker:2009hy}, the authors proposed a new Wilson loop operator that 
couples both to the bosonic and fermionic fields of the theory using explicitly the quiver 
structure of the theory. This operator turns out to be 1/2 BPS (if it lies on a straight line or circular contour) and it is dual to the 1/2 BPS 
string solution found in \cite{Drukker:2008zx,Rey:2008bh}.
There are more general BPS Wilson loops, lying along 
arbitrary curves and preserving fewer supersymmetries. These loops have been proposed in \cite{Cardinali:2012ru} and 
they generalize the straight line and the circle studied in \cite{Drukker:2009hy}. 
Clearly it is possible to consider also non-supersymmetric operators as, for 
instance, Wilson loops with cusps or selft-intersections.

In this section we review the bosonic and fermionic Wilson loop operators in ABJ(M) theory and 
their weak coupling expansion.

\subsection{The bosonic Wilson loops}\label{sec:GYloops}

In ABJM theory there are quite a few possibilities to construct 
gauge-invariant Wilson loop operators: one choice would be simply the standard Wilson loop 
operators which are defined respect to the gauge holonomies $W_A$ and $\hat{W}_A$ of the $U(N)_\kappa$ and $U(M)_{-\kappa}$ groups respectively as follows
\begin{equation}\begin{split}\label{WA}
\mathcal{W}_A[C]=&\frac{1}{N}\text{Tr}_{\mathbf{N}}(W_A[C])=\frac{1}{N}\text{Tr}_{\mathbf{N}}\mathcal{P}\exp\left[-i\int_C d\tau\,A_\mu 
\dot{x}^\mu\right]\,,\\
\hat{\mathcal{W}}_A[C]=&\frac{1}{M}\text{Tr}_{\mathbf{M}}(W_A[C])=\frac{1}{M}\text{Tr}_{\mathbf{M}}\mathcal{P}\exp\left[-i\int_C d\tau\,\hat{A}_\mu 
\dot{x}^\mu\right]\,,
\end{split}\end{equation}
where $x^\mu(\tau)$ parametrizes the curve $C$ along which the loop operator is supported, $\text{Tr}_{\mathbf{N}}$ and $\text{Tr}_{\mathbf{M}}$ represent the color traces taken over the fundamental of $U(N)_\kappa$ and $U(M)_{-\kappa}$ respectively and $\mathcal{P}$ is the path-ordering operator.

The loop operators \eqref{WA} are not BPS. In order to have an operator locally BPS at least, one has to define the connection with some coupling to the scalars. The resulting operators are
\begin{equation}\begin{split}\label{ABJMWL16}
\mathcal{W}_B[C]=&\frac{1}{N}\text{Tr}_{\mathbf{N}}(W_B[C])=\frac{1}{N}\text{Tr}_{\mathbf{N}}\,\mathcal{P}\exp\left[-i\!\!\int_C \!d\tau\!\left( A_\mu \dot{x}^\mu-\frac{2\pi i}{\kappa}
|\dot{x}|M_{J}^{\ \ I} C_I \bar{C}^J\right)\right]\,,\\
\hat{\mathcal{W}}_B[C]=&\frac{1}{M}\text{Tr}_{\mathbf{M}}(\hat{W}_B[C])=\frac{1}{M}\text{Tr}_{\mathbf{M}}\,\mathcal{P}\exp\left[-i\!\!\int_C \!d\tau\!\left( \hat{A}_\mu \dot{x}^\mu-\frac{2\pi i}{\kappa}
|\dot{x}|\hat{M}_{J}^{\ \ I} \bar{C}^J C_I\right)\right]\,,
\end{split}\end{equation}\normalsize
where $M_J^{\ \ I}$ and $\hat{M}_J^{\ \ I}$ in general are matrices with arbitrary entries but that can be constrained by supersymmetry. 

Using the supersymmetry transformation \eqref{susytransfABJ}, one can consider the variation of the operators \eqref{ABJMWL16} under supersymmetry and 
impose that it vanishes for a suitable choice of the parameter $\theta$. For instance, considering the first operator of \eqref{ABJMWL16}, one finds the following condition
\begin{equation}\begin{split}
\delta_\theta \mathcal{W}_B\sim 
\theta^{IJ\alpha}[-\dot{x}_\mu &\sigma_{\alpha\beta}^\mu\delta_I^P+|\dot{x}|\delta_{\alpha\beta}M_I^{\ \ P}]C_P(\psi_J)^\beta\\
&+\epsilon_{IJKL}\theta^{IJ\alpha}[\dot{x}_\mu\sigma_{\alpha\beta}^\mu\delta_P^K+|\dot{x}|\delta_{\alpha\beta}M_P^{\ \ K}](\bar{\psi}^L)^\beta\bar{C}^P 
=0
\end{split}\end{equation}
In order to have a supersymmetric operator, both terms in the formula above have to vanish separately.
The choice of which supercharges are preserved by the operator completely fixes 
the scalar coupling $M_J^{\ \ I}$ and the geometry of the contour $\dot{x}^\mu$ and 
vice-versa. Indeed if the contour is a straight line, the Wilson line operator is invariant under two of the 12 Poincar\'e supersymmetries and two of the 
12 superconformal supersymmetries, \textit{i.e.} the loop is 1/6 BPS\footnote{
The line can be mapped to a circle under a conformal transformation. Then the circular Wilson loop possesses the same number of supersymmetries 
of the Wilson line but, since such transformation mixes the super-Poincar\'e and
superconformal charges, it is invariant under a linear combination of them.}.

In general, one can take any combination of $\mathcal{W}_B[C]$ and $\hat{\mathcal{W}}_B[C]$  in any representation 
of each of the gauge groups. An explicit computations in planar perturbation theory \cite{Drukker:2008zx,Rey:2008bh} shows that the operator 
identifiable with the appropriate Type IIA fundamental string configurations\footnote{
The fundamental string ending along a straight line on the boundary of $AdS_4$ and localized on 
$\mathbb{CP}^3$ preserves 12 supercharges. In order to match with the gauge theory observable one has to smear the string over a $\mathbb{CP}^1$, 
breaking indeed the supersymmetry down to 1/6.} is the sum of $\mathcal{W}_B$ and $\hat{\mathcal{W}}_B$.

\subsection{The fermionic Wilson loop}\label{sec:DTloop}

The idea of \cite{Drukker:2009hy} is to define the 1/2 BPS Wilson line (or circular Wilson loop)\footnote{All the statements of this section are true for a generic contour but the operator will preserves less supercharges.} embedding the gauge connections 
of $U(N)_{\kappa}$ and $ U(M)_{-\kappa}$ into a superconnection of the form
\begin{equation}
  \label{superconnection}
 \mathcal{L}(\tau) \equiv \begin{pmatrix}
A_{\mu} \dot x^{\mu}-\frac{2 \pi i}{k} |\dot x| \mathcal{M}_{J}^{\ \ I} C_{I}\bar C^{J}
&i\sqrt{\frac{2\pi}{k}}  |\dot x | \eta_{I}\bar\psi^{I}\\
-i\sqrt{\frac{2\pi}{k}}   |\dot x | \psi_{I}\bar{\eta}^{I} &
\hat  A_{\mu} \dot x^{\mu}-\frac{2 \pi i}{k} |\dot x| \hat{\mathcal{M}}_{J}^{\ \ I} \bar C^{J} C_{I}
\end{pmatrix}
\end{equation}
belonging to the super-algebra of $U(N|M)$. 
The matrices $\mathcal{M}_{J}^{\ \ I}$ and $\hat{\mathcal{M}}_{J}^{\ \ I}$ and the Grassmann even quantities $\eta_I^\alpha$ and $\bar{\eta}^I_\alpha$ parameterize the possible local couplings. 
The above supermatrix $\mathcal{L}$ is in general  $(N+M)\times(N+M)$: the upper-left block is $N\times N$ and the 
lower-right is $M\times M$. In particular, the diagonal entries have the same structure of the gauge 
connections of the bosonic Wilson loops $\mathcal{W}_B[C]$ and $\hat{\mathcal{W}}_B[C]$ 
respectively discussed above with different scalar couplings. For a given path $C$, the holonomy of the superconnection 
\eqref{superconnection} is defined as
\begin{equation}\label{WLnontraced}
W[C]\equiv \mathcal{P} \exp{\left(-i\int_{C} 
d\tau\,\mathcal{L}(\tau)\right)}.
\end{equation}

In the AdS/CFT framework it is useful to require that the Wilson loop possesses 
a local $U(1)\times SU(3)$ R-symmetry invariance as the semi-classical string 
surface. This requirement restricts the coupling to be of the form
 \begin{equation}\label{couplingWL12}
\eta_I^\alpha=n_I 
\eta^\alpha\,,\quad\bar{\eta}^I_\alpha=\bar{n}^I\bar{\eta}_\alpha\,,\quad 
\mathcal{M}_J^{\ \ I}=p_1\delta^I_J-2p_2 n_J\bar{n}^I\,,\quad
\hat{\mathcal{M}}_J^{\ \ I}=q_1\delta^I_J-2q_2 n_J\bar{n}^I\,
\end{equation}
 where the reduced vector coupling $n_I$ (and its complex conjugate $\bar{n}^I$) define the embedding of the unbroken $SU(3)$ subgroup 
into $SU(4)$ and it can be chosen such that $n_I\bar{n}^I=1$. The free parameters $p_i(\tau)$ and $q_i(\tau)$, and the reduced spinor
couplings can be constrained by requiring that the Wilson loop defined by the superconnection 
\eqref{superconnection} is globally supersymmetric up to a super-gauge 
transformation\footnote{The usual condition $\delta_{\theta}\mathcal{L}(\tau)=0$ gives rise only to bosonic Wilson loops $(\eta=\bar\eta=0)$ and at most 1/6 BPS.}
\begin{equation}\label{BPSABJ}
\delta_{\theta}\mathcal{L}(\tau)=\mathcal{D}_\tau G\equiv\partial_\tau G+i\{\mathcal{L},G]
\end{equation}
with $G$ an anti-diagonal supermatrix. The resulting couplings are
\begin{equation}\begin{split}\label{condizioneeta}
{(\dot{x}^\mu\gamma_\mu)_\alpha}^\beta=\frac{\ell}{2i}& |\dot{x}|
(\eta^\beta\bar\eta_\alpha+\eta_\alpha\bar{\eta}^\beta)\,,\qquad
(\eta^\beta\bar\eta_\alpha+\eta_\alpha\bar{\eta}^\beta)=2 i\delta^\beta_\alpha\,,\\
&\mathcal{M}_J^{\ \ I}=\hat{\mathcal{M}}_J^{\ \ I}=\ell(\delta^I_J-2 n_J\bar{n}^I)\,,
\end{split}\end{equation}
where $\ell=\pm 1$. Notice that the vectors $n_I$ and $\bar{n}^I$ are still unconstrained. 

In order to determine the number of supersymmetries preserved by a certain fermionic Wilson 
loop, one has to solve the general set of BPS conditions given in \cite{Cardinali:2012ru} 
using the following ansatz for the reduced vector couplings
\begin{equation}
\label{bnIS2}
 \bar n^{I}=r (\eta U\bar s^{I})
\ \  \ \ \ \mathrm{and}\ \ \ \ \
 n_{I}=\frac{1}{r} (s_{I}U^{-1}\bar\eta), 
\end{equation}
where $ s_{I}^{\alpha}$  and $\bar  s^{I}_{\alpha}$ are $\tau$-independent spinors obeying the completeness relation $\bar{s}^I_\beta 
s_I^\alpha=~\tfrac{1}{2i}\delta^\alpha_\beta$, and
\begin{equation}
\label{Umatrix}
r=r_0 e^{-\frac{i\tau}{2}\sin2\alpha}\quad\text{and}\quad U=\cos\alpha~\mathbb{1}+i \sin\alpha~ (x^{\mu}\gamma_{\mu}),
\end{equation}   
with $\alpha$ free constant parameter. The resulting couplings are the following
\small\begin{equation}\label{genericdgrtabjm}
\begin{split}
\eta^{\beta}_{I}=&
\frac{i}{r_{0}}e^{\frac{i}{2}(\sin2\alpha) \tau}\left[s_{I}(\cos\alpha~\mathbb{1}-i \sin\alpha~ (x^{\mu}\gamma_{\mu}))\left(\mathbb{1}+\frac{\dot{x}\cdot \gamma}{|\dot x|}\right)\right]^{\beta},
\\
\bar\eta_{\beta}^{I}=&
i r_{0}e^{-\frac{i}{2}(\sin2\alpha) \tau}\left[\left(\mathbb{1}+\frac{\dot{x}\cdot \gamma}{|\dot x|}\right)\left(\cos\alpha~\mathbb{1}+i \sin\alpha~ (x^{\mu}\gamma_{\mu})\right)\bar s^{I}\right]_{\beta},
\\
\mathcal{M}_{K}^{\ \ J} \!=&\hat{\mathcal{M}}_{K}^{\ \ J}
\!\!=\!\left [\delta^{J}_{K}\!-\!2i s_{K}\bar s^{J}\!-\!2 i\cos 2\alpha  \!\left(\!s_{K}\frac{\dot{x}\cdot \gamma}{|\dot x|} \bar s^{J}\!\right)\!-\!2 i \sin2\alpha \left(s_{K}\gamma^{\lambda}\bar s^{J}\right)\epsilon_{\lambda\mu\nu} x^{\mu}\dot x^{\nu}
\right].
\end{split}
\end{equation}\normalsize
If the contour is parametrized in $\mathbb{R}^3$ the ansatz \eqref{bnIS2} 
has $\alpha=0$ and consequently the couplings \eqref{genericdgrtabjm}. The 
associated Wilson loops are 1/12 BPS and they are the three-dimensional 
companions of the operators discussed in \cite{Zarembo:2002an}. In this case the most 
supersymmetric operator is the 1/2 BPS Wilson line.
If the contour is parametrized on the sphere $S^2$, the parameter $\alpha$ 
is arbitrary and the associated Wilson loops are in general 1/12 BPS . 
It is interesting to mention that for $\alpha=0$ 
the scalars decouple from the 2-forms on $S^2$ (\textit{i.e.} the last term in the scalar couplings $\mathcal{M}$ and $\hat{\mathcal{M}}$)
and the corresponding Wilson loops can be interpreted as a deformation of the 
previous case. For $\alpha=\tfrac{\pi}{4}$, scalars couple only to the invariant forms and one can recover 
the 1/2 BPS circle, the 1/6 BPS latitude and the 1/6 BPS two-longitude. 
These Wilson loops can be interpreted as the three-dimensional 
companions of the operators discussed in \cite{Drukker:2007dw,Drukker:2007qr}.

Finally one has to consider the boundary 
conditions obeyed by the gauge functions (\textit{i.e.} the entries of the matrix $G$ defined in \eqref{BPSABJ}) to obtain a gauge invariant object.
If the holonomy is defined on a closed contour the Wilson loop operator is 
defined as the trace of \eqref{WLnontraced}.
For an infinite open circuit, such as the straight line, there are two 
possible gauge-invariant operators
\begin{equation}\begin{split}\label{trstr}
\mathcal{W}_{-}[C]=\frac{1}{N-M}
\mathrm{STr}\left[W[C]\right]\quad\text{and}\quad\mathcal{W}_{+}[C]=\frac{1}{N+M}
\mathrm{Tr}\left[W[C]\right].
\end{split}\end{equation}
where the trace and the supertrace are taken in the fundamental of $U(N)_{\kappa}$ 
and $U(M)_{-\kappa}$ for the $N\times N$ and $M\times M$ blocks of the 
supermatrix respectively.

\subsection{The Wilson loops weak coupling expansion}

The vacuum expectation value of a fermionic Wilson loop is given by the usual path 
integral representation
\begin{equation}\label{vevgeneral}
\langle\mathcal{W}_{\pm}[C]\rangle=\frac{1}{N\pm M}\int\mathcal{D}[\Psi]\;e^{-S_{\text{ABJ(M)}}}
\;\text{(S)Tr}[W[C]]
\end{equation}
where $\mathcal{D}[\Psi]$ specifies the integration over all the fields of the 
theory and $S_{\text{ABJ(M)}}$ is the action of the ABJ(M) theory in the Euclidean 
space (see appendix \ref{sec:appendixA2}). The symbol (S)Tr stands for the trace or the supertrace 
respectively for $\mathcal{W}_+$ and $\mathcal{W}_-$ (see \eqref{trstr}).

Since the holonomy $W[C]$ is a supermatrix and one has to take its (super)trace, 
it is useful to split the vev \eqref{vevgeneral} into the two diagonal partial vevs as 
follows
\begin{equation}\label{vevpm}
\langle\mathcal{W}_{\pm}[C]\rangle=\frac{\langle\mathcal{W}^{\uparrow}[C]\rangle\pm \langle\mathcal{W}^{\downarrow}[C]\rangle}{N\pm 
M}\, .
\end{equation}
The partial vevs $\mathcal{W}^{\uparrow(\downarrow)}[C]$ are defined by
\begin{equation}\label{vevarrow}
\langle\mathcal{W}^{\uparrow(\downarrow)}[C]\rangle=\int\mathcal{D}[\Psi]\;e^{-S_{\text{ABJ(M)}}}
\;\text{Tr}_{\textbf{N}(\textbf{M})}[W^{\uparrow(\downarrow)}[C]]\,,
\end{equation}
where the arrows $\uparrow$ and $\downarrow$ specify respectively the upper-left $N\times N$ and the
lower-right $M\times M$ blocks of the supermatrix $W[C]$. Recalling the 
definitions \eqref{superconnection} and \eqref{WLnontraced}, the path-ordered 
expansion of $W^\uparrow[C]$ is given by\footnote{The path ordered expansion of $W^\downarrow[C]$ is the same
of $W^\uparrow[C]$ with the following replacements: $N\leftrightarrow M$, $\mathcal{A}_i\leftrightarrow \hat{\mathcal{A}}_i$ and $(\eta\bar\psi)_i\leftrightarrow(\psi\bar\eta)_i$.} 
\begin{equation}\begin{split}
\label{expaloop}
\text{Tr}_{\textbf{N}}(W^\uparrow[C])=\text{Tr}_{\textbf{N}}\biggl[&\mathbb{1}_N-i\int_C d\tau_1{\mathcal A}_1-\int_{C}d\tau_{\mbox{\tiny $\displaystyle1\!\!>\!\! 2$}}\biggl({\mathcal A}_1{\mathcal A}_2+|\dot{x}_1||\dot{x}_2|(\eta\bar{\psi})_1(\psi\bar{\eta})_2 \biggr)\\
&+i\int_{C}d\tau_{\mbox{\tiny $\displaystyle1\!\!>\!\! 2\!\!>\!\!3$}}\biggl( {\cal A}_1{\cal A}_2{\cal A}_3-\frac{2\pi}{k}[|\dot{x}_1||\dot{x}_2|(\eta\bar{\psi})_1(\psi\bar{\eta})_2{\cal A}_3\\ 
&\qquad +|\dot{x}_1||\dot{x}_3|(\eta\bar{\psi})_1\hat{\cal A}_2 (\psi\bar{\eta})_3+|\dot{x}_2||\dot{x}_3|{\cal A}_1(\eta\bar{\psi})_2(\psi\bar{\eta})_3]\biggr)+...\biggr]
\end{split}\end{equation}
where the subscripts specify the dependence of the fields from the position $x_i = x(\tau_i)$ and $\mathcal{A}$ and $\hat{\mathcal{A}}$ refers to the upper-left and lower-right diagonal blocks of the superconnection \eqref{superconnection} resepctively. The spinor and 
R-symmetry indices are suppressed 
($\eta\bar{\psi}\equiv \eta_{I}^{\alpha}\bar\psi^{I}_{\alpha}$ and $\psi\bar{\eta}\equiv 
\psi_{I}^{\alpha}\bar{\eta}^{I}_{\alpha}$). The path-ordered integrals in \eqref{expaloop} 
are defined as follows
\begin{equation}\label{nestedint}
\int_C d\tau_{\mbox{\tiny $\displaystyle1\!\!>\!\! 2\!\!>\!\! ... 
\!\!>\!\!n-1\!\!>\!\!n$}}=\int_a^b d\tau_1\int_a^{\tau_1}d\tau_2 \;\,...\int ...\int_a^{\tau_{n-2}} 
\!\!d\tau_{n-1}\int_a^{\tau_{n-1}}\!\!d\tau_n
\end{equation}
for $\tau_i \in [a,b]$. An interesting property of the partial vevs defined in 
\eqref{vevarrow} is
\begin{equation}\label{symvev}
\langle\mathcal{W}^{\uparrow}[C]\rangle\overset{N\leftrightarrow M}=
\langle\mathcal{W}^{\downarrow}[C]\rangle
\end{equation}
that clearly drastically simplify the computation of \eqref{vevpm}.

The same considerations can be done for the bosonic Wilson loops. The diagonal entries of the superconnection \eqref{superconnection}
resemble the gauge connections of the bosonic Wilson loops $\mathcal{W}_B[C]$ and $\hat{\mathcal{W}}_B[C]$ defined in \eqref{ABJMWL16} but with different scalar couplings. Indeed the vev of the bosonic Wilson loops are different from the vev computed from the fermionic operator turning off the fermionic couplings (in some cases this equivalence is true, for example the circular Wilson loop at framing one\footnote{In Chern-Simons theories the vev of Wilson loop operators on close paths could be affected by finite regularization ambiguities if it is used a point-splitting regularization for contact singularities which appear when multiple points collide. In this regularization scheme this can be avoided by requiring that every point $x_i$ runs on a different path called frame (for example in pure Chern-Simons theory see \cite{Witten:1988hf,Guadagnini:1989am,Alvarez:1991sx}).} \cite{Drukker:2009hy,Drukker:2010nc}).
Since our purpose is to provide an algorithmic method to expand the Wilson loops at weak coupling, we don't need to specify the value of the scalar coupling at any time. The two bosonic Wilson loops are the trace of the holonomy of $U(N)$ and $U(M)$ and they are related to the upper-left and lower-right blocks of the superconnection \eqref{superconnection} respectively. The expansion of bosonic loop operators can be reconstructed form the fermionic one (see \eqref{expaloop}) with some simple substitutions.
Indeed, setting the fermionic coupling $\eta$ and $\bar\eta$ to zero in the expansions of $\text{Tr}_{\textbf{N}}(W^\uparrow[C])$ or $\text{Tr}_{\textbf{M}}(W^\downarrow[C])$, and substituting the scalar couplings $\mathcal{M}_J^{\;\;I}\rightarrow {M}_J^{\;\;I}$, we obtain the expansion for the bosonic loops operator as follows
\begin{equation}\begin{split}
\text{Tr}_{\mathbf{N}}(W_B[C])=&\text{Tr}_{\mathbf{N}}(W^\uparrow[C])_{\eta=\bar\eta=0,\,\mathcal{M}\rightarrow M}\\
\text{Tr}_{\mathbf{M}}(\hat{W}_B[C])=&\text{Tr}_{\mathbf{M}}(W^\downarrow[C])_{\eta=\bar\eta=0,\,\mathcal{M}\rightarrow M}
\end{split}\end{equation}
and for the operators $\mathcal{W}_A[C]$ and $\hat{\mathcal{W}}_A[C]$
\begin{equation}\begin{split}
\text{Tr}_{\mathbf{N}}(W_A[C])=&\text{Tr}_{\mathbf{N}}(W^\uparrow[C])_{\eta=\bar\eta=\mathcal{M}=\hat{\mathcal{M}}=0}\\
\text{Tr}_{\mathbf{M}}(\hat{W}_A[C])=&\text{Tr}_{\mathbf{M}}(W^\downarrow[C])_{\eta=\bar\eta=\mathcal{M}=\hat{\mathcal{M}}=0}
\end{split}\end{equation}
Then one can read the weak coupling expansion of the bosonic Wilson loops from the fermionic one defined in \eqref{expaloop}.

In ABJ(M) theory the Chern-Simons level $\kappa$ plays the role of a coupling. Indeed 
one can define the coupling constant $g^2_{CS}=\tfrac 1 \kappa$ that plays a 
similar role of $g^2_{YM}$ in $\mathcal{N}=4$ SYM, albeit $\kappa$ has 
to be an integer to preserve non-abelian gauge symmetry. Then it is possible to study the vev of the Wilson loop
with the usual perturbation theory technique at weak coupling ($g_{CS}\ll 1$) as 
follows
\begin{equation}\label{PT}
\langle\mathcal{W}^{\uparrow}[C]\rangle\equiv \sum_{\ell=0}^\infty 
g_{CS}^{2\ell}\,a_\ell^\uparrow(N,M)=\sum_{\ell=0}^\infty 
\frac{a_\ell^\uparrow(N,M)}{\kappa^\ell}
\end{equation}
where using \eqref{symvev} we have $a_\ell^\downarrow(N,M)=a_\ell^\uparrow(M,N)$. The coefficients 
$a_\ell^\uparrow$ can be expressed in terms of Feynman diagrams. For $\ell=0,1$ we have 
\begin{equation}\begin{split}\label{coeffa}
a^\uparrow_0&=N\\
a^\uparrow_1&=\vcenter{\hbox{\includegraphics[width=3.7cm]{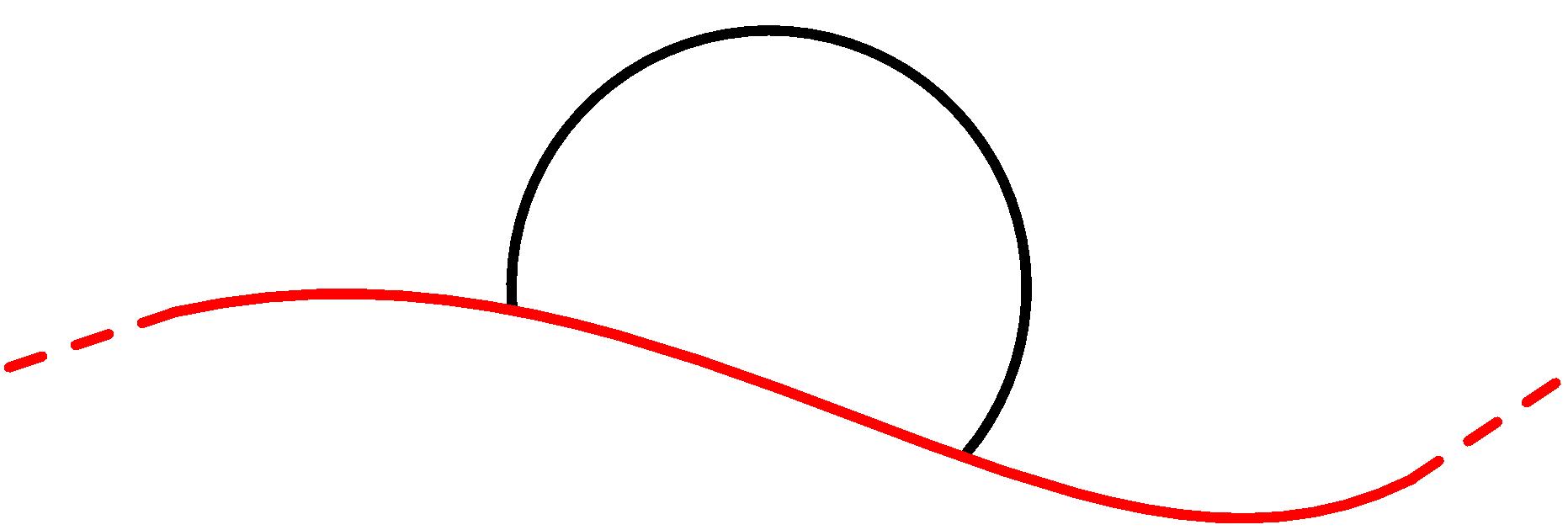}}} + \vcenter{\hbox{\includegraphics[width=3.7cm]{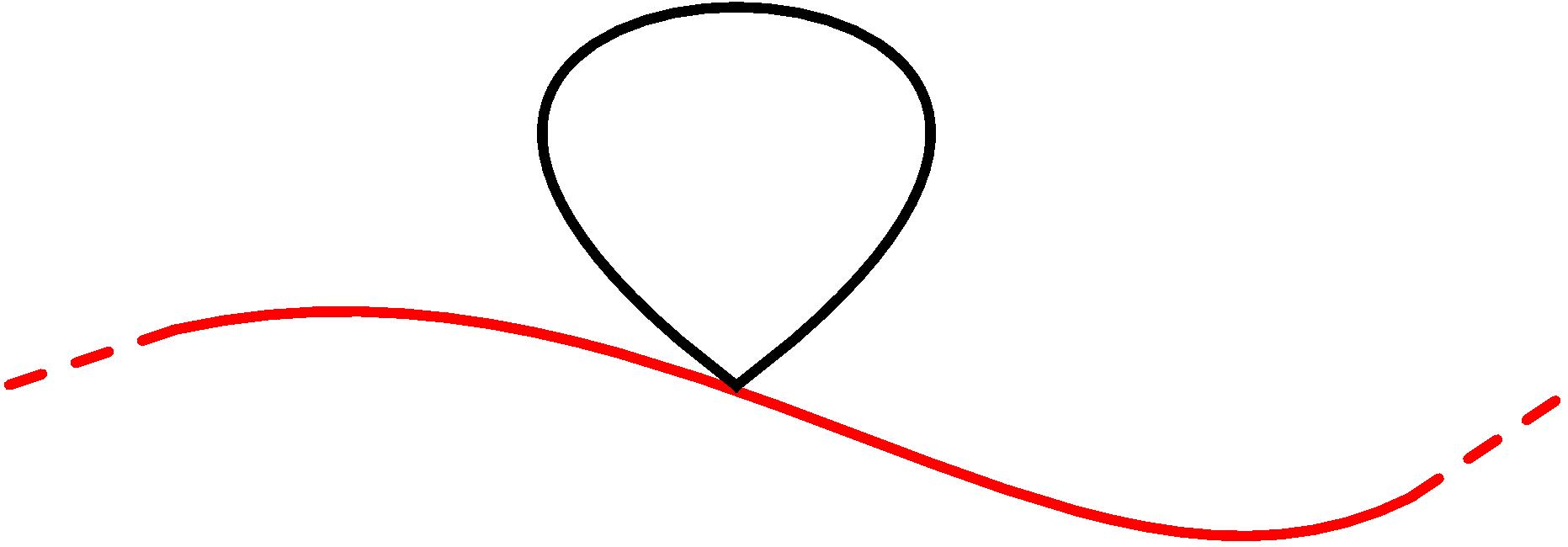}}} +\vcenter{\hbox{\includegraphics[width=3.7cm]{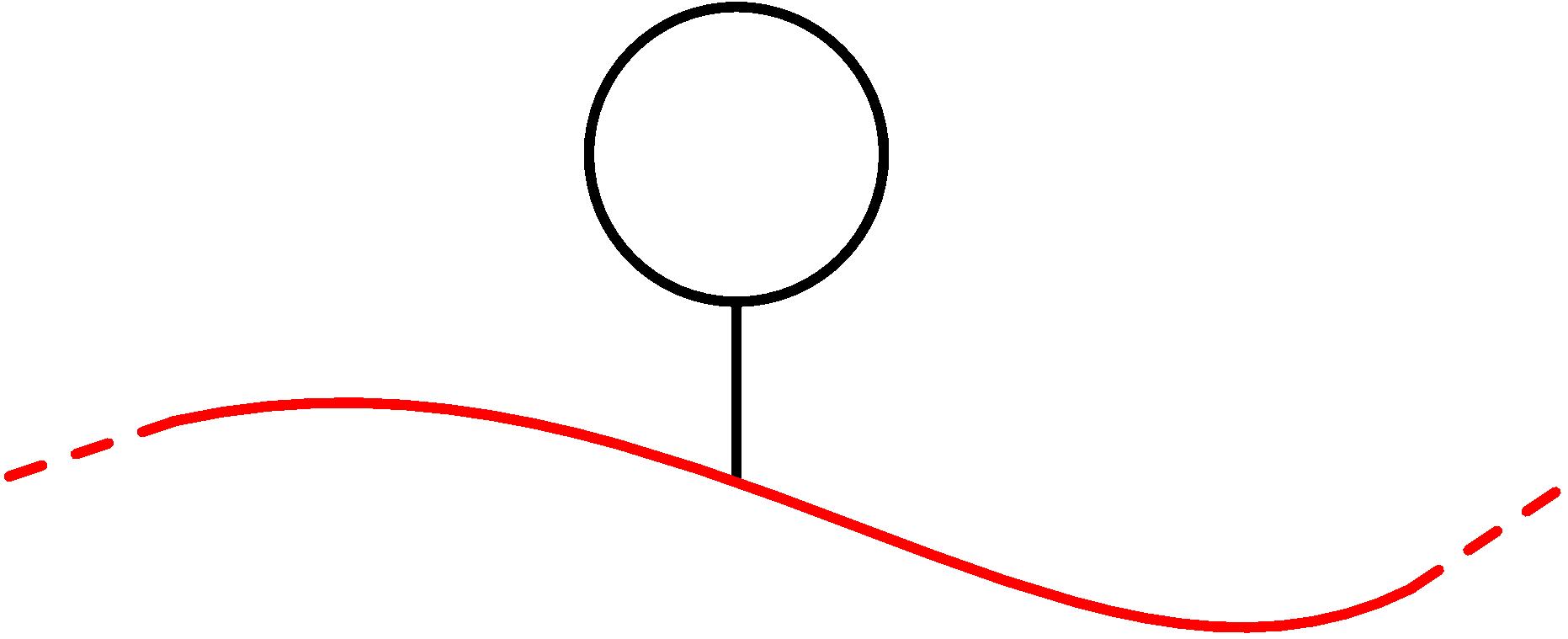}}}
\end{split}\end{equation}
where the red line represent the Wilson loop and the black lines are propagators (it is not specified the type of the propagators because 
here we are interested only in the topology of the diagrams).
Finally the observable can be restricted in the planar 't Hooft limit which introduces the following effective couplings
\begin{equation}\label{thooft}
\lambda_1\equiv g_{CS}^2 N=\frac{N}{\kappa}\,,\qquad
\lambda_2\equiv g_{CS}^2 
M=\frac{M}{\kappa}\qquad\text{with}\;\;\kappa,M,N\rightarrow\infty
\end{equation}
In the ABJM case, since $M=N$, the theory has only one coupling $\lambda=\lambda_1=\lambda_2$.

\section{The \texttt{WiLE} \textit{Mathematica}\textsuperscript{\textregistered} package}\label{sec:sec2}

In this section we introduce the package \textit{Mathematica}\textsuperscript{\textregistered} \texttt{WiLE} and its usage.
The purpose of the package is to compute the weak coupling expansion of Wilson loops in ABJ(M) in terms of Feynman diagrams as in \eqref{PT} for any contour and at any loop order specified by the user. 
In the following we present the setup of the package and a detailed manual.

\subsection{Download and installation}

The package set up is very simple. All the relevant files can be downloaded from 
the arXiv servers with the source files of this paper\footnote{In the abstract page of this paper click on ``Other formats'' and then on ``Download source''. 
The downloaded file is a compressed directory without extension that needs to be uncompressed. The package and a \textit{Mathematica}\textsuperscript{\textregistered}
notebook containing an explicit example can be found in the ``WiLE'' directory.}. The most updated version of the package can be also downloaded from the GitHub repository at the following address

 \url{https://github.com/miciosca/WiLE}.

The package \texttt{WiLE.m} can be loaded into any \textit{Mathematica}\textsuperscript{\textregistered} 
notebook saved in the same directory with the command 

\vspace{.1cm}
\noindent\hspace{.5cm}\includegraphics[scale=.8]{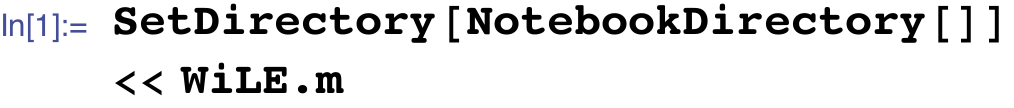}
\vspace{.1cm}

\noindent or

\vspace{.1cm}
\noindent\hspace{.5cm}\includegraphics[scale=.8]{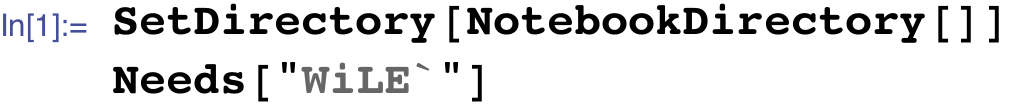}
\vspace{.1cm}

It is also possible to install the package in such a way that it is available 
for any notebook without specify any directory. To do it, select ``File'', 
then ``Install'' and then follow the instructions. After installing, the package 
can be loaded with the commands above without the function 
\texttt{SetDirectory}.

\subsection{Manual}\label{sec:manual}

The new available functions can be listed using the command

\vspace{.1cm}
\noindent\hspace{.5cm}\includegraphics[scale=.8]{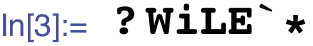}
\vspace{.1cm}

and they are the following
\begin{itemize}
  \item \texttt{WiLE}: The main function of the package. Given the number and the 
  type of the fields on the Wilson loop and the vertices from the action, the 
  function draws all the possible diagrams with the related integrands;
  \item \texttt{WiLEFullorder}: It draws diagrams with the related integrands for 
  a given order in the weak coupling expansion of the Wilson loop;
  \item \texttt{WiLESimplify}: It simplify the integrands. 
\end{itemize}
In the following we briefly describe their applications. 

\paragraph{WiLE}: When the package is loaded, run the command 

\vspace{.1cm}
\noindent\hspace{.5cm}\includegraphics[scale=.8]{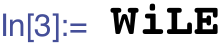}
\vspace{.1cm}

without semicolon at the end. The output will be the following window
\begin{flalign}\label{WiLEpanel}
&\hspace{.5cm}\includegraphics[scale=.5]{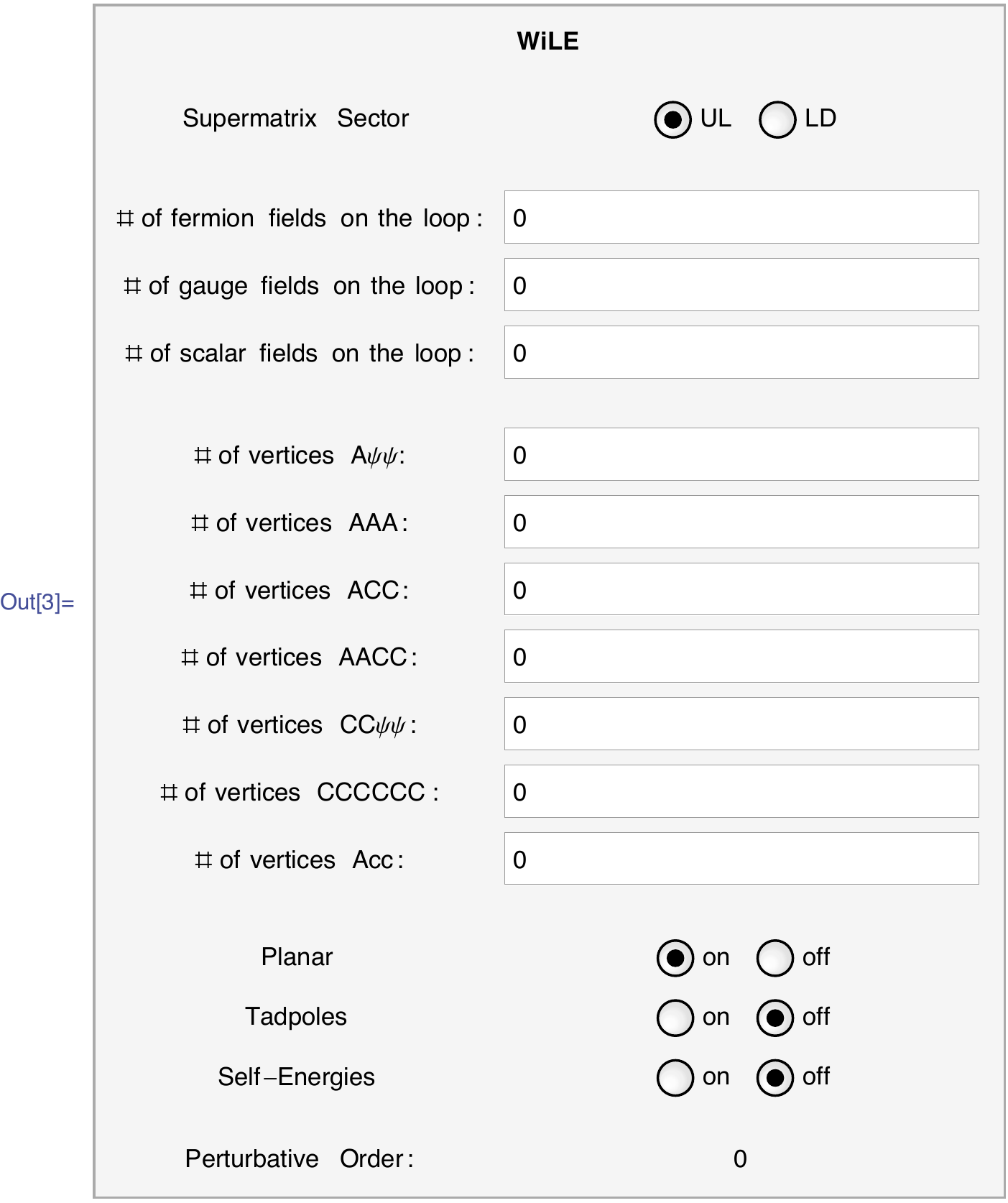}\hspace{.5cm}\includegraphics[scale=.1]{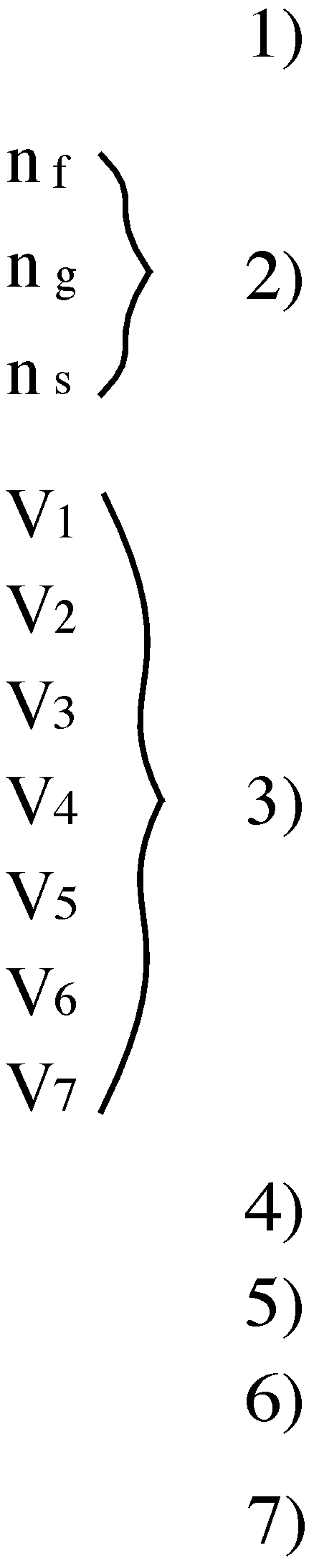}&
\end{flalign}
The \texttt{WiLE} output can be computed filling the input fields with a suitable set of initial data and selecting the 
wanted options and then pressing the usual \texttt{Shift+Enter}. In the following we have 
listed the instructions for any input line in \eqref{WiLEpanel}\footnote{The list numbers are related to the numbers appearing in the panel \eqref{WiLEpanel}. These labels are added for simplicity and then they are not present in the \texttt{WiLE} input window.}. 
\begin{enumerate}
  \item \textbf{Supermatrix Sector}: This option allows the user to choose 
  between the upper-left or the lower right blocks of the supermatrix $W[C]$. 
  In other words, choosing the options \texttt{UL} or \texttt{LR} the diagrams in the output will be 
  part of the vev's $\langle \mathcal{W}^\uparrow[C]\rangle$ or $\langle \mathcal{W}^\downarrow[C]\rangle$ 
  respectively (see \eqref{vevarrow});
  \item \textbf{\# of fermion, gauge and scalar fields on the loop}: The user 
  has to specify the number (a positive integer or 0) of the three available fields on the Wilson loop: $n_f$ fermion fields $\psi$ or $\bar\psi$,
  $n_g$ gauge fields $A_\mu$ or $\hat{A}_\mu$ and $n_s$ couples of scalar fields $C\bar C$ or $\bar C C$. The above 
 \texttt{Supermatrix sector} option is the discriminant between the two possibilities for any $n_i$. 
  Given the triplet $(n_f,\,n_g,\,n_s)$, the algorithm selects the wanted monomials in the expansion of
  $\text{Tr}_{\mathbf{N}(\mathbf{M})}(W^{\uparrow(\downarrow)}[C])$. To give an example, for the
  path-ordered expansion of $\text{Tr}_{\mathbf{N}}(W^{\uparrow}[C])$ in 
  \eqref{expaloop}, the triplet $(0,\,0,\,0)$ selects the first term 
  $\mathbb{1}_{\mathbf{N}}$, the triplet $(0,\,1,\,1)$ selects the $\mathcal{A}_1\mathcal{A}_2$ 
  monomial in which both the connections can contribute as a gauge field or a couple of scalars (see \eqref{superconnection}) but not the same simultaneously and the triplet
  $(2,\,1,\,0)$ selects the last three monomials in which $\mathcal{A}$ 
  contributes as a gauge field. It is clear from the equation \eqref{expaloop}, 
  that the choice of the triplet $(n_f,\,n_g,\,n_s)$ fixes the number of nested 
  integrals \eqref{nestedint}. Indeed if the output of 
  \texttt{WiLE} for a given triplet $(n_f,\,n_g,\,n_s)$ is $F(n_f,\,n_g,\,n_s)$ 
  the related integral is
  \begin{equation}\label{intx}
    \int_C d\tau_{\mbox{\tiny $\displaystyle1\!\!>\!\! 2\!\!>\!\! ... 
\!\!>\!\!n-1\!\!>\!\!n$}}\,F(n_f,\,n_g,\,n_s)\qquad\text{with}\quad 
n=n_f+n_g+n_s
  \end{equation}
  \item \textbf{\# of vertices}: The user has to specify the number (a positive integer or 0)
  of the seven available interaction vertices of the action. More in detail:
  \begin{itemize}
    \item \textbf{\# of vertices $A\psi\psi$}: It is the number $V_1$ of 
    three-gauge-fermion vertices coming from the term $\bar{\psi}^I\slashed{D}\psi_I$ of the matter action $S_{\text{matter}}$ (see 
    \eqref{Smatter});
    \item \textbf{\# of vertices $AAA$}: It is the number $V_2$ of three-gauge vertices 
    coming from the terms $A_\mu A_\nu A_\rho$ and $\hat{A}_\mu \hat{A}_\nu \hat{A}_\rho$ 
    of the Chern-Simons action $S_{\text{CS}}$ (see \eqref{Scs});
    \item \textbf{\# of vertices $ACC$}: It is the number $V_3$ of 
    three-gluon-scalar vertices coming from the term $D_\mu C^I D^\mu\bar{C}_I$ of the matter action  $S_{\text{matter}}$ (see 
    \eqref{Smatter});
    \item \textbf{\# of vertices $AACC$}: It is the number $V_4$ of 
    four-gluon-scalar vertices coming from the term $D_\mu C^I D^\mu\bar{C}_I$ of the matter action  $S_{\text{matter}}$ (see 
    \eqref{Smatter});
    \item \textbf{\# of vertices $CC\psi\psi$}: It is the number $V_5$ of 
    four-fermion-scalar vertices coming from the Yukawa term $S_{\text{pot}}^F$ of the action (see 
    \eqref{SF});
    \item \textbf{\# of vertices $CCCCCC$}: It is the number $V_6$ of 
    six-scalar vertices coming from the sextic potential term $S_{\text{pot}}^B$ of the action (see 
    \eqref{SB});
    \item \textbf{\# of vertices $Acc$}: It is the number $V_7$ of 
    three-gauge-ghost vertices coming from the terms $\partial_\mu\bar{c}D^\mu c$ and
    $\partial_\mu\bar{\hat{c}}D^\mu\hat{c}$ of the gauge-fixing action $S_{\text{gf}}$ (see 
    \eqref{Sgf}).
  \end{itemize}
  According to our notation, the position of the vertices is denoted by the letter $z_i$ with the index 
  $i=1,...,v$ where $v$ is the total number of vertices $v=\sum_{i=1}^7V_i$.
  It is clear from the definition \eqref{vevarrow} that any vertex comes with an 
  integration over $z$ over all the space $\mathbb{R}^3$. Then, if the output of 
  \texttt{WiLE} for a given number of vertices is $F(V_1,V_2,V_3,V_4,V_5,V_6,V_7)$ 
  the related integral is
  \begin{equation}\label{intz}
  \int\! d^3z_1\!\!\int d^3z_2...\int ...\int\! d^3z_v 
  \,F(V_1,V_2,V_3,V_4,V_5,V_6,V_7)\qquad\text{with}\quad v=\sum_{i=1}^7V_i
  \end{equation}
  We set the dimension of the above integrals to be $D=3$ but it is possible to 
  keep the dimension arbitrary to dimensional-regularize the integrals (for more details see the explicit example in section \ref{sec:Example}). 
  \item \textbf{Planar}: This option allows the user to choose to expand the 
  Wilson loop in the planar 't Hooft limit \eqref{thooft} or not. Since the color structure of a given diagram in general is  
  $M^sN^t$ with $s,t=0,1,...$, the quantity $s+t$ has to obey to the 
  following bound
  \begin{equation}\label{bound}
  0\leq s+t\leq p-v+1
  \end{equation}
  with $p$ the number of propagators. If the option \texttt{Planar} is \texttt{on}, then the 
  algorithm selects only the planar diagrams, namely the diagrams that saturate 
  the upper bound of \eqref{bound}. Otherwise the output will contain all the 
  possible color structures for any diagram.
  \item \textbf{Tadpoles}: This option allows the user to choose if tadpoles diagrams are listed into 
  the output or not. From a topological point of view, tadpoles are identified as diagrams
  that are not connected anymore if one of the vertices is deleted (here we are considering the position on the loop $x_i$ as vertex).
  For instance the last two diagrams in \eqref{coeffa} are tadpoles. Their value depends on the regularization scheme:
  in dimensional regularization is well known that this kind of diagrams vanish then, if this is the selected scheme, it is useful to select the option \texttt{Tadpoles} on \texttt{off}.
  In general, for a given set of initial data, the number of tadpole diagrams is huge, then selecting \texttt{off} the number of diagrams in the output will be significantly reduced.
  \item \textbf{Self-Energies}: This option allows the user to choose if self-energy diagrams are listed into 
  the output or not. From a topological point of view, self-energies are defined as diagrams
  that are not connected anymore if two edges are cut. From a physical 
  point of view, self-energies are the diagrams in which one or more of the 
  propagators are substituted by a $\ell$-loop corrected propagator with $\ell$ depending on the perturbative order.
  For instance, at one-loop there are no self-energies (see \eqref{coeffa}), at 
  two-loop the self-energies can be computed starting from the one-loop diagrams 
  and dressing them with the one-loop effective propagators and so on. Using this method, the number of loop integrations decrease and then it is useful to select the option \texttt{Self-Energies} on \texttt{off} and reads the self-energies from the dressing of the previous perturbation theory orders.
  Otherwise, since the $\ell$-loop effective propagators are not known at any $\ell$, the algorithm will print the 
  self-energy diagrams as a explicit $\ell$-loop integrals.    
  \item \textbf{Perturbative order}: Given the number of the fields on the loop $(n_f,\, n_g,\, n_s)$ 
  and the number of vertices $V_i$ with $i=1,...,7$, it shows in real-time 
  the perturbation theory order $\ell=p-v$ in which the diagrams in the output are 
  included. It is clear from the definition \eqref{PT} that $\ell$ has to be a positive 
  integer number. If it is not, the message \texttt{$\notin$~Integers} will be displayed. 
\end{enumerate}
The output of the function \texttt{WiLE} is a matrix. Any row of the matrix 
consists of two elements: the first is the diagram drawing and the second is the 
related integrand. For instance, if one sets the number of fields on the loop as 
$(n_f=2,\, n_g=1,\, n_s=1)$, the number of vertices as $V_1=V_5=1$ and the other to zero and 
the remaining options as in \eqref{WiLEpanel}, the function will generate a 
matrix with 24 rows. Since the perturbative order is $\ell=4$, these 24 diagrams contribute to 
$\tfrac{a_4^\uparrow(N,M)}{\kappa^4}$ (see \eqref{PT}) then they are involved in the four-loop computation of
$\langle\mathcal{W}_{\pm}[C]\rangle$ (see \eqref{vevpm}). One of these diagrams for example is the following 
\vspace{-.3cm}
\begin{center}\noindent\includegraphics[trim=1.1cm 0 0 0, clip=true, scale=.4]{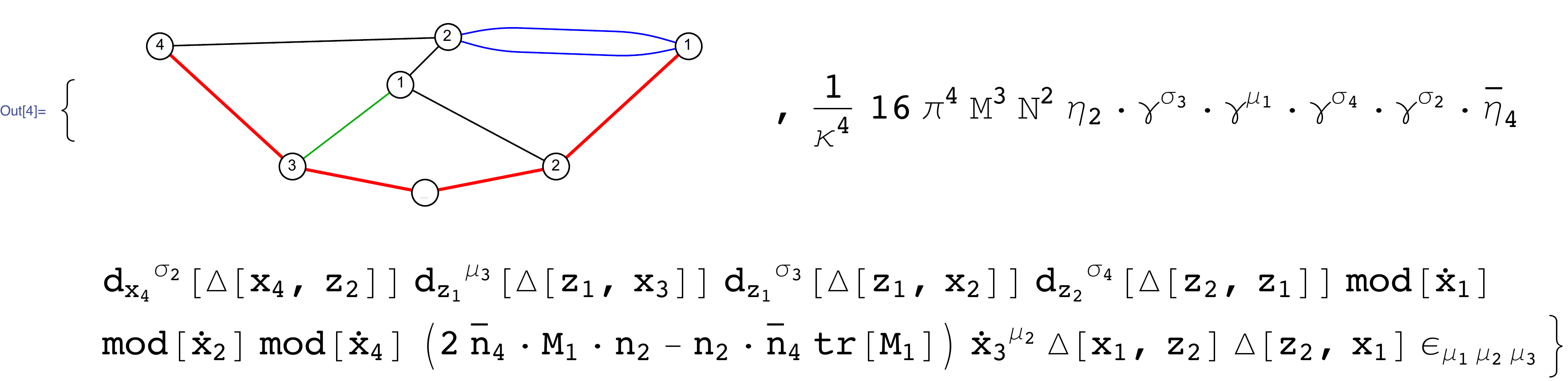}\end{center}
\vspace{-.3cm}
where the integrand is written using the built-in \textit{Mathematica}\textsuperscript{\textregistered} package 
\texttt{Notation}. The symbols appearing in the 
integrand are written in order to be as much as possible similar to the related 
quantities defined in section \ref{sec:susyWLABJM}. The complete set of symbols involved in the output is listed in appendix \ref{sec:appendixB}. Briefly, the red thick 
lines represents the Wilson loop, the black lines are fermions propagators, 
the green line is a gauge propagator and the blue lines are scalar 
propagators. The encircled numbers on top and outside the Wilson loop stand for the 
positions $x_i$ with $i=1,...,4$ and $z_j$ with $j=1,2$ respectively. The loop is oriented 
counter-clockwise then $\tau_1>\tau_2>\tau_3>\tau_4$. It is possible to recover 
the number and the kind of the integrations in front of this output simply looking at the 
diagram. Indeed, following the prescription given in \eqref{intx} and \eqref{intz} and 
using the legend in appendix \ref{sec:appendixB}, we can write 
\begin{equation}\begin{split}
&\vcenter{\hbox{\includegraphics[trim=1.1cm 0 0 0, clip=true, scale=.28]{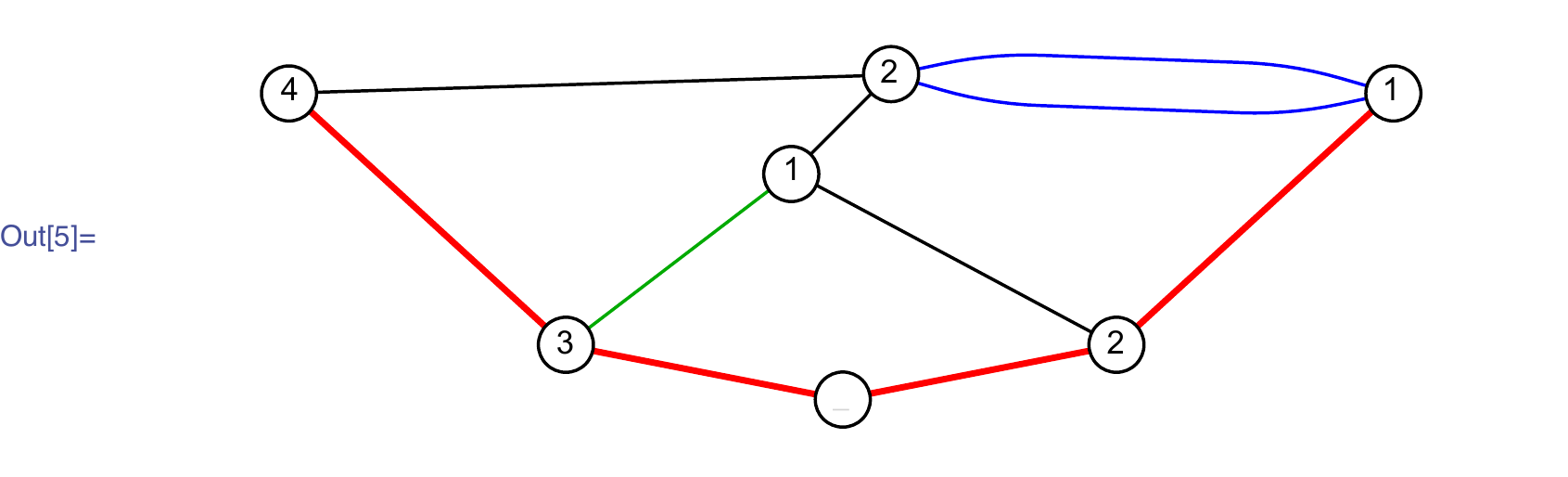}}}\!\!=\left(\frac{2\pi}{\kappa}\right)^4\!\!M^3N^2\!\! \int_C \!d\tau_{\mbox{\tiny $\displaystyle1\!\!>\!\! 2\!\!>\!\! 3 
 \!\!>\!\!4$}}\!\int d^3z_1d^3z_2\biggl( 
|\dot{x}_1||\dot{x}_2||\dot{x}_4| \\
&\qquad\qquad\qquad\qquad\times[2(\bar{n}_4\mathcal{M}_1n_2)-({n}_2\bar{n}_4)\text{tr}(\mathcal{M}_1)]\epsilon_{\mu_1\mu_3\nu_3}\dot{x}_3^{\nu_3}(\eta_2\gamma^{\sigma_3}\gamma^{\mu_1}\gamma^{\sigma_4}\gamma^{\sigma_2}\bar{\eta}_4)\\
&\qquad\qquad\qquad\qquad\times\Delta(x_1,z_2)^2\partial_{x_2}^{\sigma_3}\Delta(x_2,z_1)
\partial_{x_3}^{\mu_3}\Delta(x_3,z_1)\partial_{x_4}^{\sigma_2}\Delta(x_4,z_2)
\partial_{z_1}^{\sigma_4}\Delta(z_1,z_2)\biggr)
\end{split}\end{equation}
where the $\partial_{x}^{\mu}=\tfrac{\partial}{\partial x^\mu}$ and $\Delta(x,y)$ is the kinematic part of the propagator defined in \eqref{propdelta}.

For a given set of initial data, the output of \texttt{WiLE} could be zero or even non existent. In 
both cases the printed output will be an empty matrix \texttt{$\{\}$}. The user will be notified by warnings
if the output is zero for some very specific reasons. For instance if the number 
of fields is odd, if all the diagrams in the output are not connected or if the 
initial data are invalid (\text{i.e.} the inputs are not zeroes or positive 
integers).

\paragraph{WiLEFullorder}: When the package is loaded, run the command 

\vspace{.1cm}
\noindent\hspace{.5cm}\includegraphics[scale=.8]{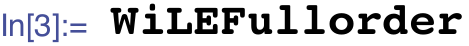}
\vspace{.1cm}

without semicolon at the end. The output will be the following window
\begin{flalign}\label{WiLEFullorderpanel}
&\hspace{.5cm}\includegraphics[scale=.5]{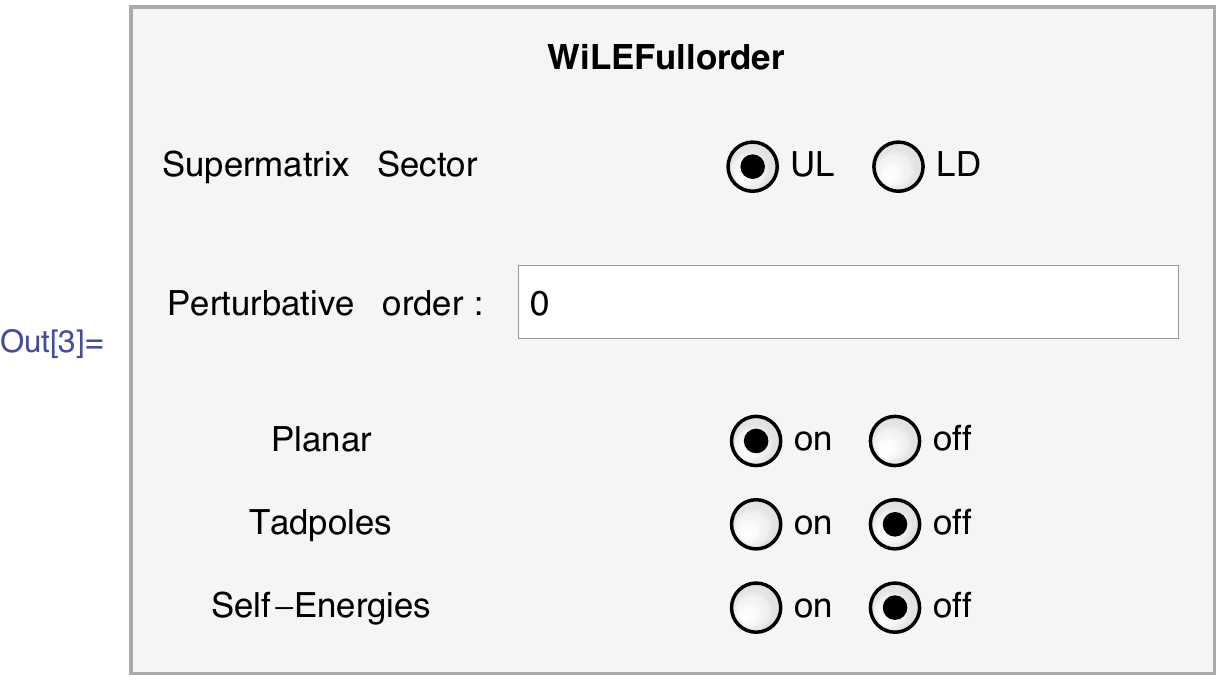}&
\end{flalign}
This function prints the complete set of diagrams and integrand for a given perturbative order $\ell$ contributing to $\tfrac{a_\ell^{\uparrow(\downarrow)}(N,M)}{\kappa^\ell}$ (see \eqref{PT}).
The input window \eqref{WiLEFullorderpanel} can be used as the previous one, the only difference is the 
kind of initial data. Indeed here the user, in addition to the options described in the previous paragraph, needs only to specify the wanted 
perturbative order $\ell$.
As the function \texttt{WiLE}, the output is a matrix with one diagram and its related integrand for any rows.
The output for $\ell=0$ is clearly the gauge group rank depending on the choice
of the supermatrix sector. The first non-trivial order is given by setting the perturbative order $\ell=1$ as we have seen schematically in \eqref{coeffa}.
In that case the output is 
\vspace{-.4cm}\begin{center}
\noindent\hspace{.5cm}\includegraphics[trim=3cm 0 6cm .4cm, clip=true, scale=.23]{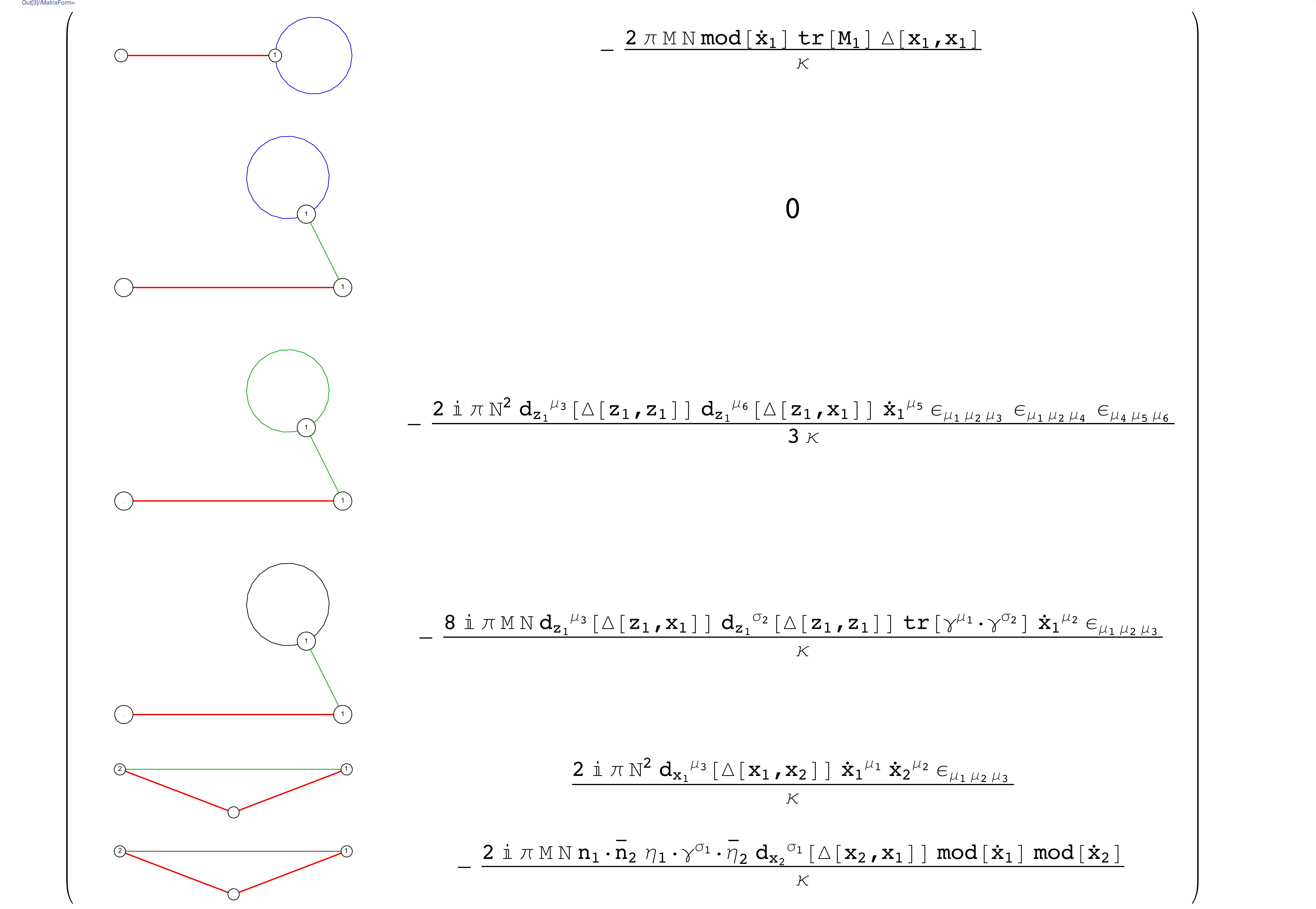}
\vspace{-.3cm}\end{center}
where we have used the built-in command \texttt{MatrixForm} for clearness 
purposes. The output above is computed selecting the most general options 
(\texttt{Planar=off}, \texttt{Tadpoles=on} and \texttt{Self-enegies=on}) then 
it contains all the possible diagrams contributing to $\tfrac{a_1^\uparrow(N,M)}{\kappa}$.
As we have seen from the equation \eqref{coeffa}, we have three different 
diagram topologies. 
Comparing this result with the topologies in \eqref{coeffa}, and adding the 
needed integrals in front of the integrands following the prescriptions given 
in \eqref{intx} and \eqref{intz}, we have
\begingroup\makeatletter\def\f@size{9.5}\check@mathfonts
\begin{align}
\label{ladder1}
\frac{1}{\kappa}\;\vcenter{\hbox{\includegraphics[width=2.5cm]{exch}}}&=\vcenter{\hbox{\includegraphics[trim=2.5cm 0 1cm 0, clip=true, scale=.32]{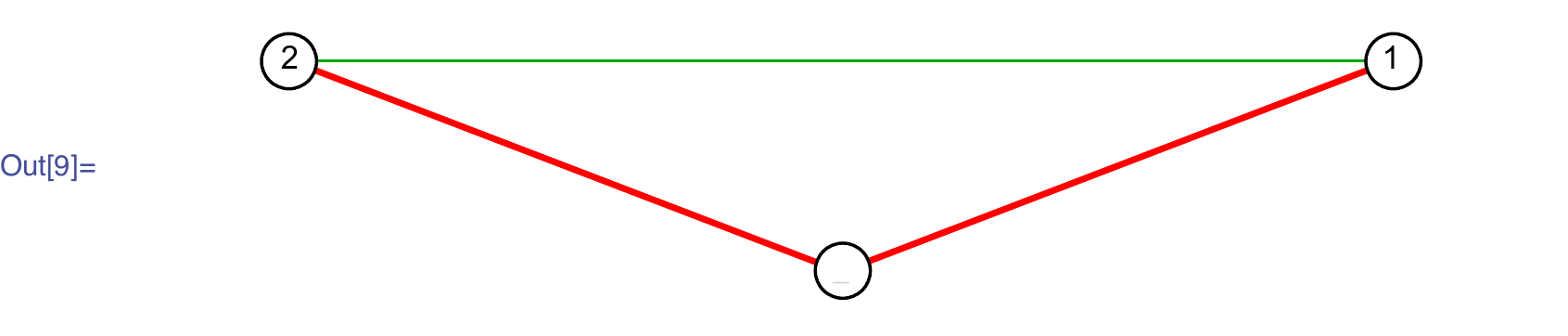}}}+\vcenter{\hbox{\includegraphics[trim=2.5cm 0 1cm 0, clip=true, scale=.32]{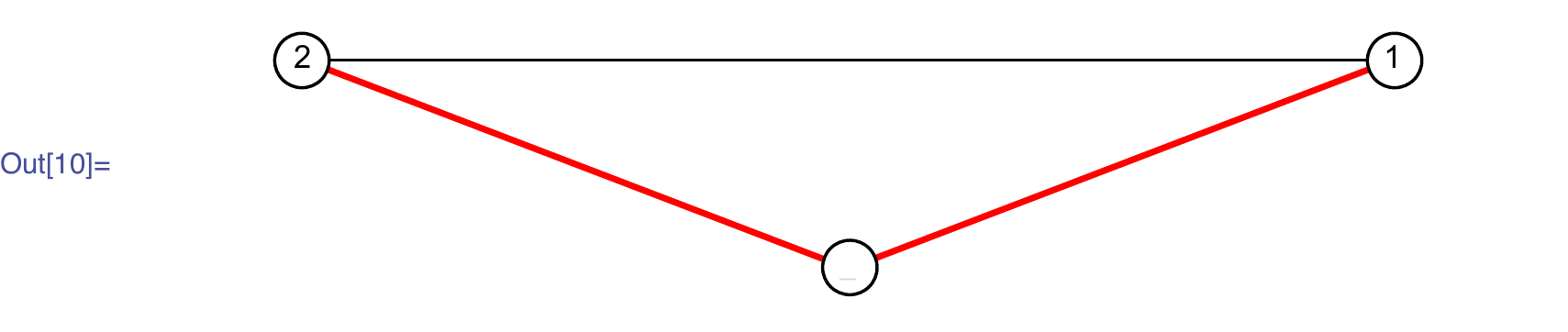}}}\nonumber\\
&\!\!\!\!\!\!\!\!\!\!\!\!\!\!\!\!\!\!\!\!\!\!\!\!\!\!\!\!\!\!\!\!\!\!\!\!\!\!\!\!\!\!\!\!=\frac{2\pi i}{\kappa}\int d\tau_{\mbox{\tiny $\displaystyle1\!\!>\!\! 2$}} 
\biggl[N^2\epsilon_{\mu_1 \mu_2 \mu_3}\dot{x}_1^{\mu_1}\dot{x}_2^{\mu_2}\partial_{x_1}^{\mu_3}\Delta(x_1,x_2)-M N |\dot{x}_1||\dot{x}_2|(n_1\bar{n}_2)(\eta_1\gamma^{\sigma_1}\bar{\eta}_2)\partial_{x_2}^{\sigma_1}\Delta(x_2,x_1)\biggr]\\
\label{tadpole1}
\frac{1}{\kappa}\;\vcenter{\hbox{\includegraphics[width=2.5cm]{tad1}}}&=\vcenter{\hbox{\includegraphics[trim=2.5cm 0 1cm 0, clip=true, scale=.29]{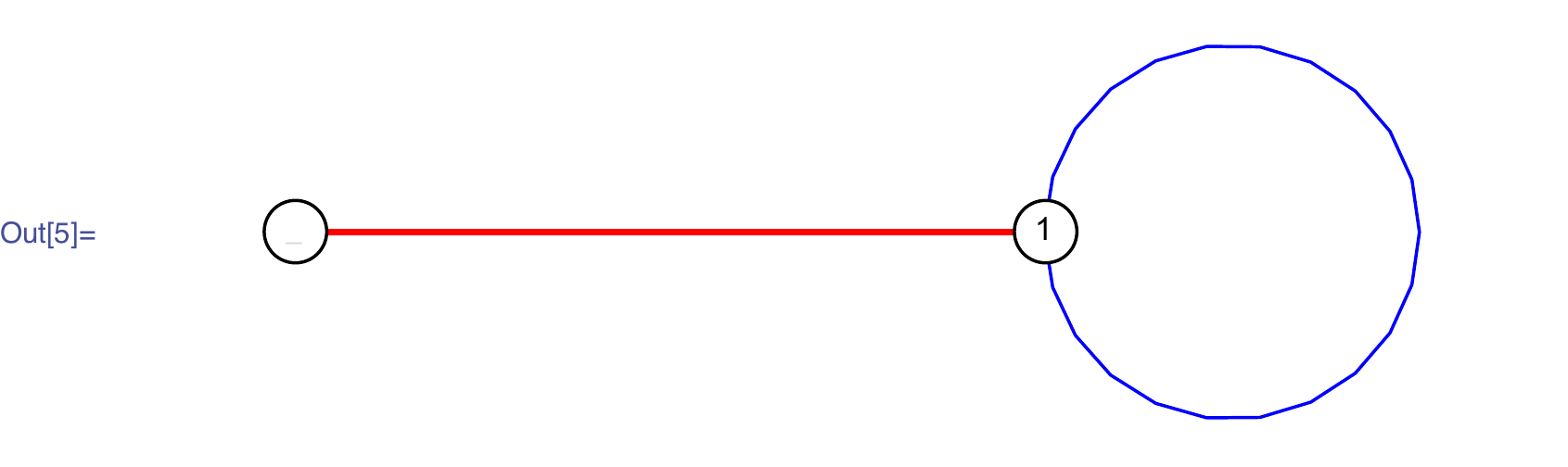}}}=-\frac{2\pi}{\kappa}M N\int d\tau_1 |\dot{x}_1|\text{tr}(\mathcal{M}_1)\Delta(x_1,x_1)\\
\label{tadpole2}\frac{1}{\kappa}\;\vcenter{\hbox{\includegraphics[width=2.5cm]{tad2}}}&=\vcenter{\hbox{\includegraphics[trim=2.5cm 0 1cm 0, clip=true, scale=.24]{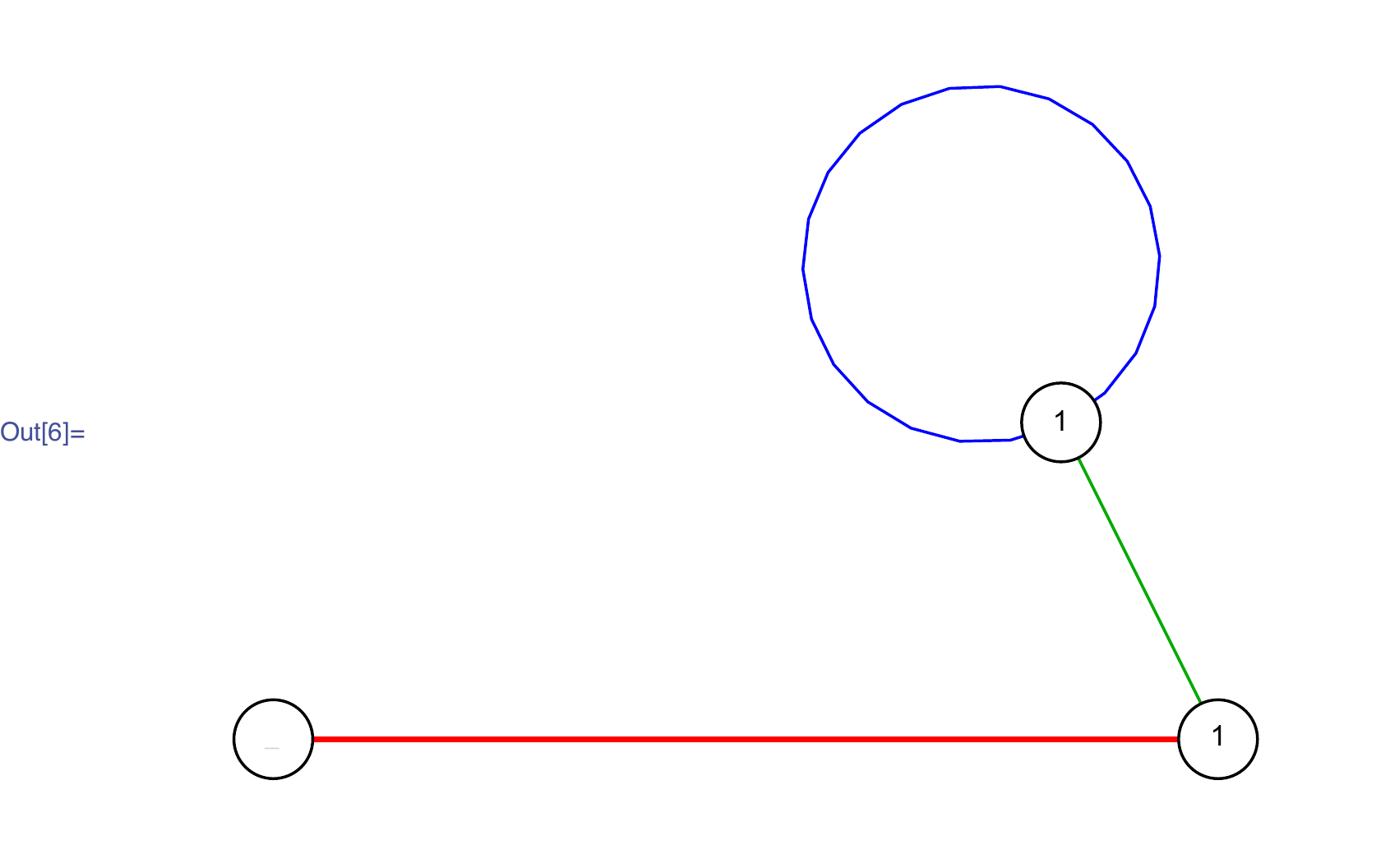}}}+\vcenter{\hbox{\includegraphics[trim=2.5cm 0 1cm 0, clip=true, scale=.24]{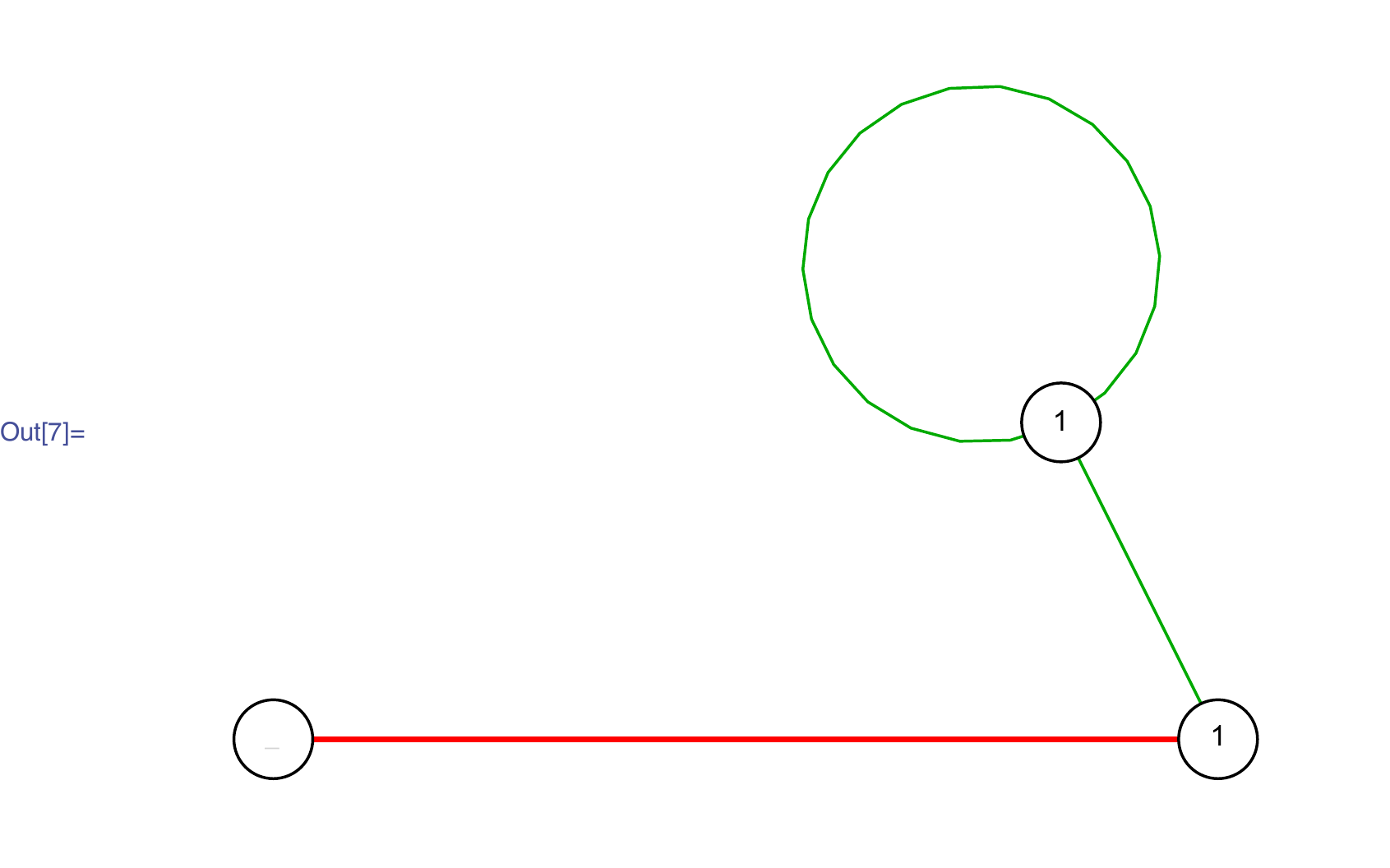}}}+\vcenter{\hbox{\includegraphics[trim=2.5cm 0 1cm 0, clip=true, scale=.24]{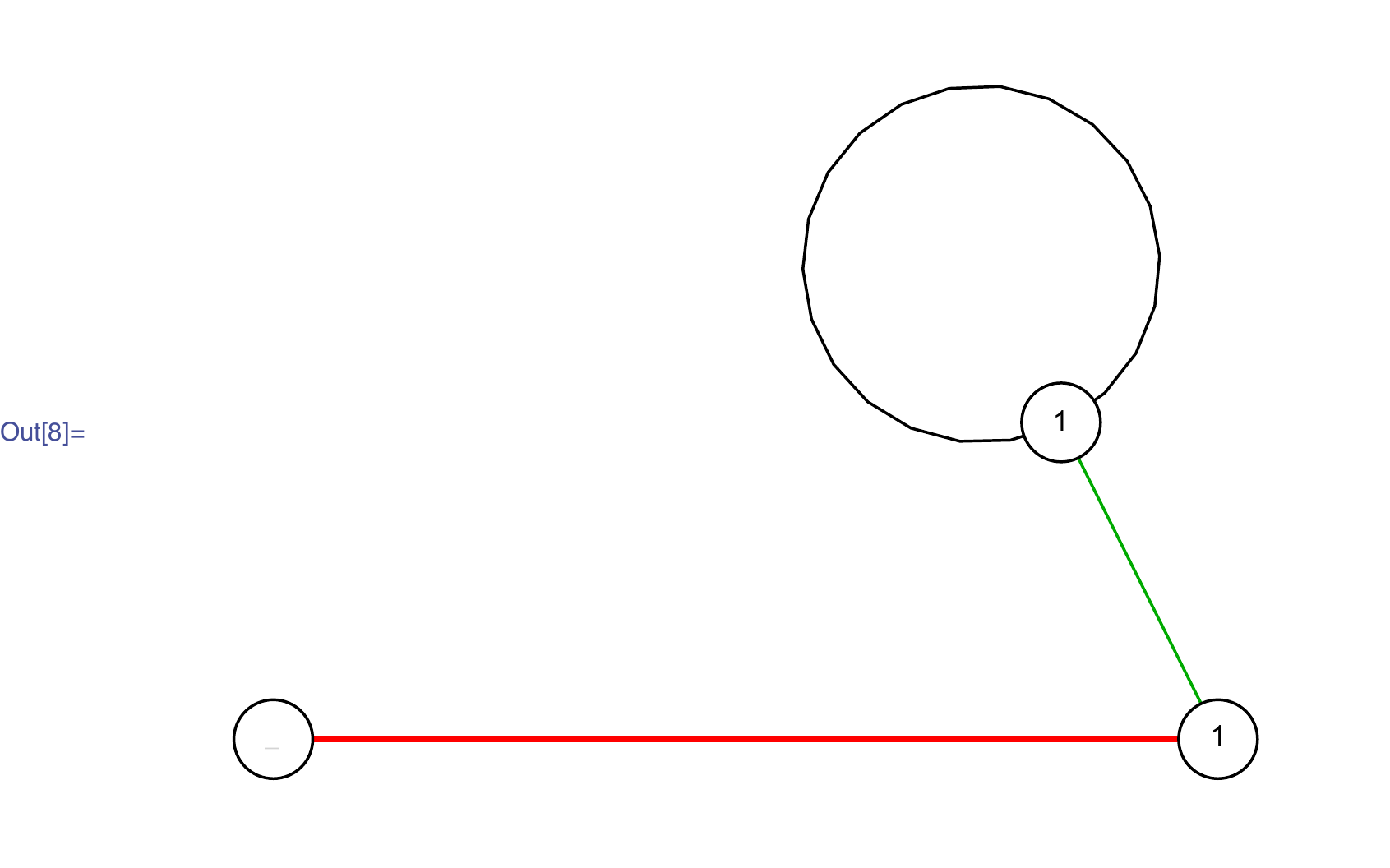}}}\nonumber\\
&\!\!\!\!\!\!\!\!\!\!\!\!\!\!\!\!\!\!\!\!\!\!\!\!\!\!\!\!\!\!\!\!\!\!\!\!\!\!\!\!\!\!\!\!=-\frac{2\pi i}{\kappa}\!\int\!\!d\tau_1\!\!\int \!\!d^3 z_1 
\biggl[\frac{N^2}{3}\epsilon_{\mu_1 \mu_2 \mu_3}\epsilon_{\mu_1 \mu_2 \mu_4}\epsilon_{\mu_4 \mu_5 \mu_6}\dot{x}_1^{\mu_5}\partial_{z_1}^{\mu_3}\Delta(z_1,z_1)\partial_{z_1}^{\mu_6}\Delta(z_1,x_1)\nonumber\\
&\qquad\qquad\qquad\quad+4M N \epsilon_{\mu_1 \mu_2 \mu_3}\text{tr}(\gamma^{\mu_1}\gamma^{\sigma_2})\dot{x}_1^{\mu_2}\partial_{z_1}^{\mu_3}\Delta(z_1,x_1)\partial_{z_1}^{\sigma_2}\Delta(z_1,z_1)\biggr]
\end{align}\endgroup
where we have used the legend in appendix \ref{sec:appendixB}.

If the input data are valid (\textit{i.e.} \texttt{Perturbative order} zero or a positive integer), the output of \texttt{WiLEFullorder} cannot be zero. The execution speed of this function decreases quickly at the raising of the perturbative order because the number of the diagrams factorially (or exponentially in the planar case) grows\footnote{More quantitatively with a laptop with 2.6 GHz Intel Core i5 processor and 8Gb or DDR3 RAM, the $\ell=0,1$ cases are computed almost instantly, $\ell=2$ in the order of seconds and $\ell=3$ in the order of minutes (the most general case needs more than 1 hour). Very high-loop cases are hard to achieve using \texttt{WiLEFullorder}, indeed we did a test for the $\ell=4$ case and the planar case needs more or less 1 day to be computed. Anyway at higher loop is easier to use the command \texttt{WiLE} for the single graph topology and parallelize the computation in order to compute many topologies at the same time. This option is not available in the current version of the package but we will implement it in a future release.}. In the following table we show the number of diagrams for a given perturbative order $\ell$ varying the options available in the input window \eqref{WiLEFullorderpanel}
\begin{center}\begin{tabular}{l *{4}{c} r}
             & $\ell=0$ & $\ell=1$ & $\ell=2$ & $\ell=3$  \\
\hline
\{\texttt{on},\texttt{off},\texttt{off}\}            & 0 & 2 & 15 & 219   \\
\{\texttt{off},\texttt{off},\texttt{off}\}            & 0 & 2 & 16 & 276   \\
\{\texttt{off},\texttt{off},\texttt{on}\}            & 0 & 2 & 24 & 826  \\
\{\texttt{off},\texttt{on},\texttt{on}\}            & 0 & 6 & 92 & 2583   \\
\end{tabular} \end{center}
where the vectors in the first column refer to $\{\texttt{Panar},\texttt{Tadpoles},\texttt{Self-Energies}\}$.

\paragraph{WiLESimplify}: This command can be used as the \textit{Mathematica}\textsuperscript{\textregistered} built-in function 
\texttt{Simplify}. The main purpose of the function \texttt{WiLESimplify} is to 
simplify expressions with a large number of $\epsilon$-tensors (both with Lorentz and R-symmetry 
indices). It is useful also for simplify very long expressions.

\section{Example: The 3-loop ladder diagrams of a Wilson line with a cusp}\label{sec:Example}

Ladder diagrams are usually the most simple class of diagrams in the 
perturbative series because they do not involve interaction vertices. 
At one-loop for instance, the total contribution to the expectation value of an arbitrary Wilson loop is given only 
by the ladder diagrams \eqref{ladder1}\footnote{If we neglect the tadpoles \eqref{tadpole1} and 
\eqref{tadpole2}.}, but at higher loops diagrams with interaction vertices start 
to contribute. In the literature there are many examples in which ladder 
diagrams play a fundamental role in the computation of some observables of 
interest. For instance the expectation value of Wilson loops on the sphere $S^2$ 
in $\mathcal{N}=4$ SYM is given only by the resummation of ladder diagrams \cite{Erickson:2000af}. The same for 
correlation functions of Wilson loops and local operators on $S^2$ \cite{Bonini:2014vta}. In \cite{Correa:2012nk} the authors identified a scaling limit in which only ladder diagrams contribute to the expectation value of Wilson loop
with a cusp. These ladders can be resummed solving a Schrodinger problem. It was also done in ABJ(M)
theory \cite{Bonini:2016fnc}. In this case the related Schrodinger problem is exactly solvable 
and it provides the solution for $\mathcal{W}_\pm$ in a closed exponential form.
The expectation value of the cusped Wilson line is divergent and its divergence is related to the generalized cusp anomalous dimension
$\Gamma_{cusp}(\lambda,\theta,\varphi)$. At $\theta=\pm\varphi$ the cusped Wilson line becomes BPS and then its anomalous dimension vanishes. Around these BPS points $\Gamma_{cusp}$ can be expanded and the first non-trivial
order is given by the Bremsstrahlung function. This function was recently 
studied up to three-loops in \cite{Bianchi:2017svd}. In order to compute this observable, the three-loop ladder diagrams was computed using the HQET formalism. 

In this section we want to present an explicit example in which the package \texttt{WiLE} is useful giving 
also a three-loop perturbative check of the exponentiation of ladders \cite{Bonini:2016fnc} 
and the ladder diagrams computed with the HQET formalism in \cite{Bianchi:2017svd}.

\subsection{The generalized cusped Wilson line}

\begin{figure}[!t]
  \begin{center}
  \includegraphics[width=10.5cm]{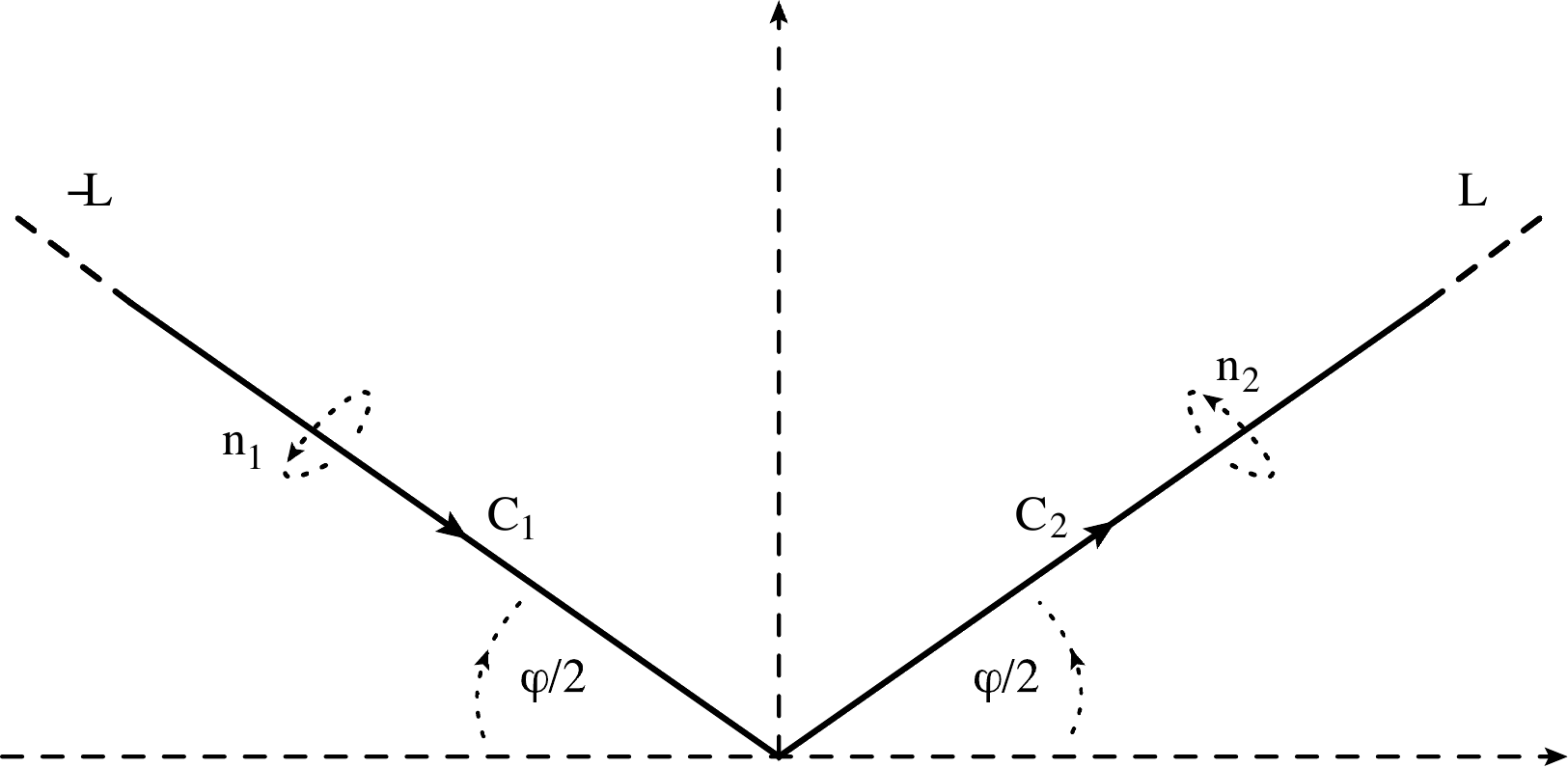}
  \caption{The planar Euclidean cusp with angular extension given by $\pi-\varphi$.}
  \label{fig:cusp}
  \end{center}
\end{figure}

Consider two Wilson lines $C_1$ and $C_2$ on $\mathbb{R}^3$ intersecting in the origin forming the curve $C=C_1\cup C_2$ as in 
Figure \ref{fig:cusp}. The angle between the two lines is $\pi-\varphi$ such that 
at $\varphi=0$ the contour becomes a straight line without the cusp. The parametrization of the contour is
\begin{equation}\label{paramcusp}
x_\mu=\{0,\tau \cos\frac\varphi 2,|\tau | \sin\frac\varphi 2\}\qquad -L\geq\tau\leq L
\end{equation} 
with $L$ an IR cut-off shielding the infinite length of the lines.

The fermionic Wilson loop operators lying on this contour are defined in 
\eqref{trstr}. In general one can choose different R-symmetry couplings for the two edges.
In other words, one can consider an angle $\theta$ that denotes the angular separation of the two Wilson lines
in the internal R-symmetry space (or $\mathbb{CP}^3$ from the string side). 
The fermionic couplings have the factorized structure presented in \eqref{couplingWL12} and they can be read from \eqref{genericdgrtabjm} setting $\alpha=0$. Since the reduced vector couplings $n$ and $\bar{n}$ are unconstrained, they can be chosen to be function of the $\theta$ angle such that
\begin{equation}
(n_1\bar{n}_2)=\cos\frac{\theta}{2}\qquad\text{and}\qquad (n_1\bar{n}_1)=(n_2\bar{n}_2)=1
\end{equation}
where the indices $1$ and $2$ specify the edges of the cusp.
In particular for the first edge they are given by
\begin{equation}\label{coupl1}
n_{1I}=\mbox{\small$ \left(\cos\frac{\theta}{4}\ \ \sin\frac{\theta}{4}\ \ 0\ \ 0\right)$}\ \ \ \ \
\eta_{1}^{\alpha}= ( e^{-i\frac{\varphi}{4}}\ \ \  e^{i\frac{\varphi}{4}})\ \ \ \ \ 
\bar n_{1}^{I}=\mbox{\footnotesize $
\begin{pmatrix}\cos\frac{\theta}{4}\\ \sin\frac{\theta}{4}\\ 0\\ 0\end{pmatrix}$} \ \ \ \ \
\bar\eta_{1\alpha} =i\begin{pmatrix} e^{i\frac{\varphi}{4}}\\ e^{-i\frac{\varphi}{4}}\end{pmatrix}
\end{equation}
and for the second edge are given by
\begin{equation}\label{coupl2}
n_{2I}=\mbox{\small$ \left(\cos\frac{\theta}{4}\ \ -\sin\frac{\theta}{4}\ \ 0\ \ 0\right)$}\ \ \ \ 
\eta_{2}^{\alpha}= ( e^{i\frac{\varphi}{4}}\ \ \  e^{-i\frac{\varphi}{4}})\ \ \ \ 
\bar n_{2}^{I}=\mbox{\footnotesize $
\begin{pmatrix}\cos\frac{\theta}{4}\\ -\sin\frac{\theta}{4}\\ 0\\ 0\end{pmatrix}$}\ \ \ \  
\bar\eta_{2\alpha} =i \begin{pmatrix} e^{-i\frac{\varphi}{4}}\\ e^{i\frac{\varphi}{4}}\end{pmatrix}
\end{equation}
From the relation \eqref{condizioneeta} one can read also the scalar couplings $\mathcal{M}$ and $\hat{\mathcal{M}}$.
Indeed using the definitions \eqref{coupl1} and \eqref{coupl2} we have
 \begin{equation}\label{coupl3}
\mathcal{M}_{1J}^{\ \ I}=
\hat{\mathcal{M}}_{1J}^{\ \ I}=\mbox{\small $\left(
\begin{array}{cccc}
 -\cos \frac{\theta }{2}& -\sin \frac{\theta }{2} & 0 & 0 \\
 -\sin \frac{\theta }{2}& \cos\frac{\theta }{2} & 0 & 0 \\
 0 & 0 & 1 & 0 \\
 0 & 0 & 0 & 1
\end{array}
\right)$}\ \ \ \ \mathrm{and}\ \ \ \  \mathcal{M}_{2J}^{\ \ I}=
\hat{\mathcal{M}}_{2J}^{\ \ I}=\mbox{\small $\left(
\begin{array}{cccc}
 -\cos \frac{\theta }{2} & \sin \frac{\theta }{2} & 0 & 0 \\
 \sin \frac{\theta }{2} & \cos\frac{\theta }{2} & 0 & 0 \\
 0 & 0 & 1 & 0 \\
 0 & 0 & 0 & 1
\end{array}
\right)$}
\end{equation}
In general the cusped Wilson line is not a BPS operator but if $\theta=\pm\varphi$, it preserves two supersymmetries. Then in this configuration the operator is globally 1/6 BPS.

The bosonic Wilson loops $\mathcal{W}_B$ and $\hat{\mathcal{W}}_B$ lying on the contour represented in figure \ref{fig:cusp} have the following bosonic couplings
 \begin{equation}\label{couplB}
M_{1J}^{\ \ I}=
\hat{M}_{1J}^{\ \ I}=\mbox{\small $\left(
\begin{array}{cccc}
 -\cos \frac{\theta }{2}& -\sin \frac{\theta }{2} & 0 & 0 \\
 -\sin \frac{\theta }{2}& \cos\frac{\theta }{2} & 0 & 0 \\
 0 & 0 & -1 & 0 \\
 0 & 0 & 0 & 1
\end{array}
\right)$}\ \ \ \ \mathrm{and}\ \ \ \  M_{2J}^{\ \ I}=
\hat{M}_{2J}^{\ \ I}=\mbox{\small $\left(
\begin{array}{cccc}
 -\cos \frac{\theta }{2} & \sin \frac{\theta }{2} & 0 & 0 \\
 \sin \frac{\theta }{2} & \cos\frac{\theta }{2} & 0 & 0 \\
 0 & 0 & -1 & 0 \\
 0 & 0 & 0 & 1
\end{array}
\right)$}
\end{equation}
Also in this case the cusped bosonic Wilson line is not a BPS operator but if $\theta=\pm\varphi$, it preserves one supersymmetry. Then in this configuration the operator is globally 1/12 BPS.

\subsection{The perturbative computation}
The evaluation of ladder diagrams of the cusped Wilson loop obviously encounters UV 
divergences which originate when the propagators endpoints coincide. 
To tame these divergences we have to use dimensional regularization. In particular we will
follow the DRED scheme in which the dimension of space-time is set to $d=3-2\epsilon$ but the dimension of $\epsilon_{\mu\nu\rho}$ tensors and the Dirac algebra
remains three. We have to introduce a mass scale $\mu^{2\epsilon}$
that keeps the action dimensionless breaking explicitly the conformal invariance.
However, because of the underlying conformal invariance, the mass scale $\mu$ 
with the IR cut-off $L$ introduced in \eqref{paramcusp} forms the combination $(\mu L)^{2\ell 
\epsilon}$ that can be always scaled away in any Feynman integral at 
$\ell$-loops.

Ladder diagrams of a fermionic Wilson loop in ABJ(M) involve three kind of propagators:
the fermionic propagator, the gauge propagator and a couple of scalar 
propagators sharing the same starting and ending points (they appear on the loop always as
the bilinears $C\bar C$ or $\bar C C$). Ladder diagrams of bosonic Wilson loops 
instead do not involve fermionic propagators.
When the contour of the loop is parametrized as a two-dimensional curve (as for instance in our case \eqref{paramcusp}),
ladder diagrams involving at least one gauge propagator vanish. Indeed, the gauge propagator 
is proportional to the following combination
\begin{equation}
\vcenter{\hbox{\includegraphics[width=4.2cm]{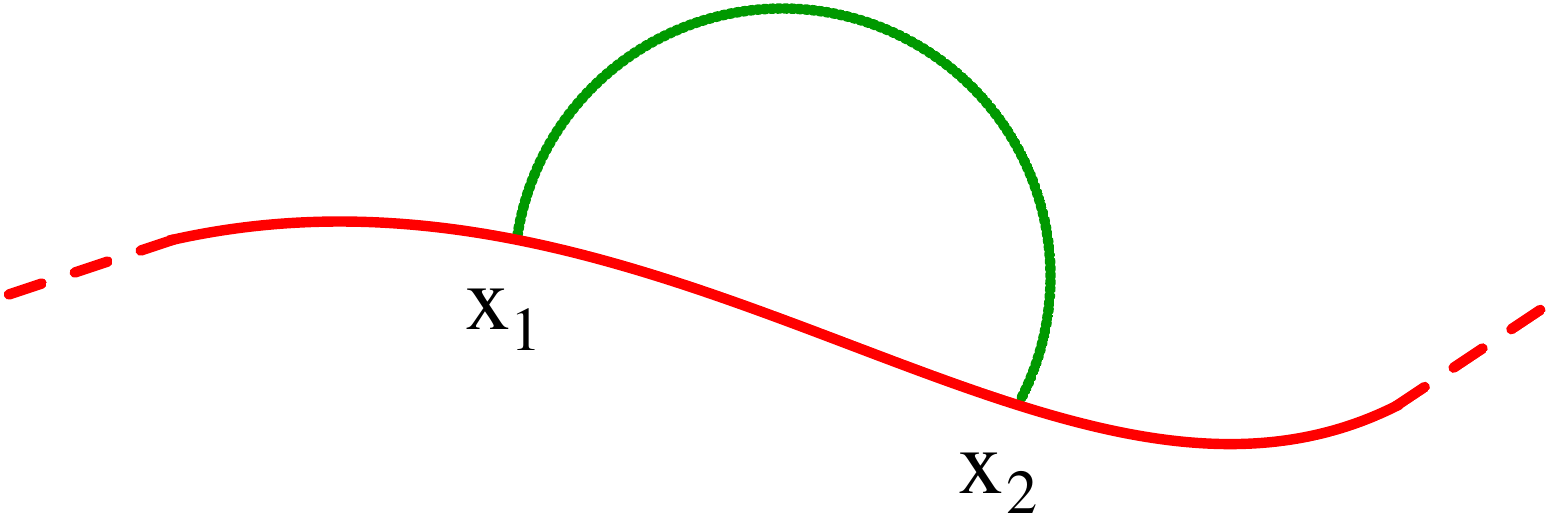}}}\propto
\epsilon_{\mu\nu\rho}\,\dot{x}_1^\mu\,\dot{x}_2^\nu\,
(x_1-x_2)^\rho=0.
\end{equation}
that is clearly zero for a two-dimensional curve.
As a consequence of this observation, since at three-loops operators $\mathcal{W}_A$ and $\hat{\mathcal{W}}_A$ contain only gauge propagators and operators $\mathcal{W}_B$ and $\hat{\mathcal{W}}_B$
contain at least one gauge propagator, the ladder part of their expectation values vanishes.
In general this is true for any odd perturbative order $\ell$, then   
\begin{equation}\begin{split}
\langle \mathcal{W}_A[C_{2d}] \rangle_{ladder}^{(\ell)}=\langle \hat{\mathcal{W}}_A[C_{2d}] 
\rangle_{ladder}^{(\ell)}&=0\qquad \ell \in\text{Odd}(\mathbb{Z}^+)\\
\langle \mathcal{W}_B[C_{2d}] \rangle_{ladder}^{(\ell)}=\langle \hat{\mathcal{W}}_B[C_{2d}] 
\rangle_{ladder}^{(\ell)}&=0\qquad \ell \in\text{Odd}(\mathbb{Z}^+)\,.
\end{split}\end{equation}
where $C_{2d}$ denotes an arbitrary two-dimensional contour.

For the fermionic Wilson loop the story is different. Indeed, at any loop order 
there is at least a class of ladder diagrams without gauge propagators (see for instance the one-loop expansion 
\eqref{ladder1}). In particular, at three-loops the ladder diagrams are given by 
the following fermionic monomial
\begin{equation}\label{t1}
-\left(\frac{2\pi}{\kappa}\right)^3|\dot{x}_1||\dot{x}_2||\dot{x}_3||\dot{x}_4||\dot{x}_5||\dot{x}_6|\text{Tr}_N\biggl[(\eta\bar{\psi})_1({\psi}\bar{\eta})_2(\eta\bar{\psi})_3({\psi}\bar{\eta})_4(\eta\bar{\psi})_5({\psi}\bar{\eta})_6\biggl]
\end{equation}
and the following mixed bosonic-fermionic monomials
\begingroup\makeatletter\def\f@size{10}\check@mathfonts
\begin{align}\label{t2}
-\left(\frac{2\pi}{\kappa}\right)^3\!\!|\dot{x}_1||\dot{x}_2|&|\dot{x}_3||\dot{x}_4|\text{Tr}_N
\biggl[(\eta\bar{\psi})_1({\psi}\bar{\eta})_2(\mathcal{M}C\bar{C})_3(\mathcal{M}C\bar{C})_4+
(\mathcal{M}C\bar{C})_1(\eta\bar{\psi})_2({\psi}\bar{\eta})_3(\mathcal{M}C\bar{C})_4\nonumber\\
&\qquad+(\eta\bar{\psi})_1(\hat{\mathcal{M}}\bar{C}C)_2(\hat{\mathcal{M}}\bar{C}C)_3({\psi}\bar{\eta})_4+
(\mathcal{M}C\bar{C})_1(\mathcal{M}C\bar{C})_2(\eta\bar{\psi})_3({\psi}\bar{\eta})_4\nonumber\\
&\qquad+(\eta\bar{\psi})_1(\hat{\mathcal{M}}\bar{C}C)_2({\psi}\bar{\eta})_3(\mathcal{M}C\bar{C})_4+
(\mathcal{M}C\bar{C})_1(\eta\bar{\psi})_2(\hat{\mathcal{M}}\bar{C}C)_3({\psi}\bar{\eta})_4\biggl]
\end{align}\endgroup
of the expansion \eqref{expaloop} where the short hand notation $(\mathcal{M}C\bar{C})_i$ and $(\hat{\mathcal{M}}\bar{C}C)_i$ stands for
$(\mathcal{M}_i)_J^{\ \ I}C_I(x_i)\bar{C}^J(x_i)$ and $(\hat{\mathcal{M}}_i)_J^{\ \ I}\bar{C}^J(x_i)C_I(x_i)$ respectively. 
Notice that we are focusing our attention on the upper-left $N\times N$ block of the supermatrix 
\eqref{WLnontraced} in order to compute $\langle\mathcal{W}^\uparrow\rangle_{ladder}^{(3)}$. The expectation value  
$\langle\mathcal{W}_\pm\rangle_{ladder}^{(3)}$ can be computed using the relation 
\eqref{symvev}. We are also restricting the computation in the large $N$ and $M$ limit 
\eqref{thooft} in which only the planar diagrams survive. For this motivation 
we do not consider some non-planar Wick contractions of the monomial \eqref{t1} and the last two 
monomials of \eqref{t2}.
Finally, we will restrict our computation for $\phi=0$. In other words we will 
compute ladder diagrams for an Euclidean Wilson line with a cusp in the 
internal R-symmetry space.
It is also useful to introduce the deformed coupling
\begin{equation}
g(\epsilon)\equiv\frac{1}{\kappa}\frac{\Gamma(\tfrac12-\epsilon)}{\pi^{\tfrac12-\epsilon}}(\mu 
L)^{2\epsilon}
\end{equation}
to simplify further the results.

In the following we will present some example of fermionic and bosonic ladder 
computation. The complete and detailed step-by-step evaluation of all the diagrams can be
found in the \textit{Mathematica}\textsuperscript{\textregistered} notebook 
``Ladders.nb'' attached to the source files of this paper. The notebook needs 
the package \texttt{HypExp} \cite{Huber:2007dx} for the Hypergeometric function expansions. It takes 
more or less 15 minutes to run. However all the computation outputs are printed 
then they can be consulted without running the entire notebook.

\paragraph{Fermionic ladder diagrams}$\\$

This class of diagrams can be computed performing the planar Wick contractions 
of the fermionic monomial \eqref{t1}. In order to do it using the package \texttt{WiLE} we have to set the
initial data to
\begin{equation}
(n_f=6,n_g=0,n_s=0)\qquad\text{and}\qquad V_i=0\quad \text{with}\quad 
i=1,2,...,7\,.
\end{equation} 
We are focusing on the upper-left part of the supermatrix then we have to select \texttt{UL}
for the option \texttt{Supermatrix Sector} and, since we want to 
compute diagrams in the 't Hooft coupling, we have to select \texttt{on} for the option 
\texttt{Planar}. The remaining available options \texttt{Tadpoles} and \texttt{Self-Energies} 
are clearly irrelevant for ladder diagrams, then we can select \texttt{off} (the algorithm is slightly quicker
choosing this option).
Given all the initial data the \texttt{WiLE} output is the following

\vspace{.1cm}
\noindent\hspace{.5cm}\includegraphics[trim=1.6cm 0 0 0, clip=true, scale=.31]{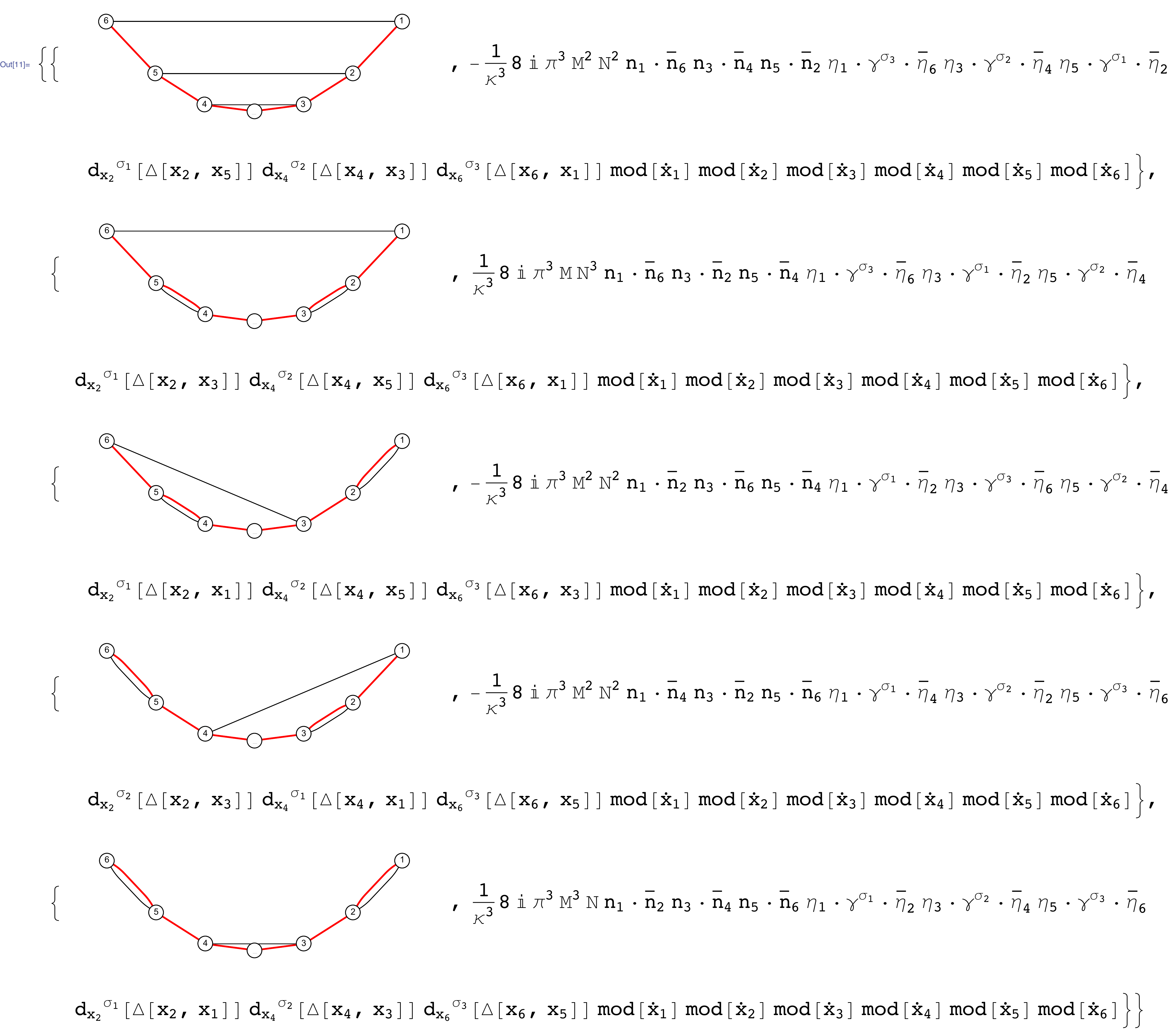}
\vspace{.1cm} 

In order to simplify the following discussion we name the diagrams above $L^\uparrow_i$ $i=1,...,5$ 
where the index $i$ corresponds to the raws of the \texttt{WiLE} output.
Since the contour is the union of two curve $C_1$ and $C_2$ and the couplings to 
the matter are different on the two edges (see \eqref{coupl1}, \eqref{coupl2} and \eqref{coupl3}), 
we have to split the region of integration for each diagram 
in seven sectors corresponding to all the possible positions of the origin. 
Luckily we do not have to compute all of them. In fact graphs which are related by a 
reflection with respect to the axis passing through the origin and orthogonal to the 
Wilson line yield the same result. Indeed using this property the number of unique subdiagrams 
for any graph $L^\uparrow_i$ decreases and we have also
\begin{equation}
L^\uparrow_4=L^\uparrow_3\,.
\end{equation}
We will compute in detail the diagram $L_1$ as an example.
\begin{figure}[!t]
 \centering
  \subfigure[$L_1^1$]
   {\includegraphics[width=3.2cm]{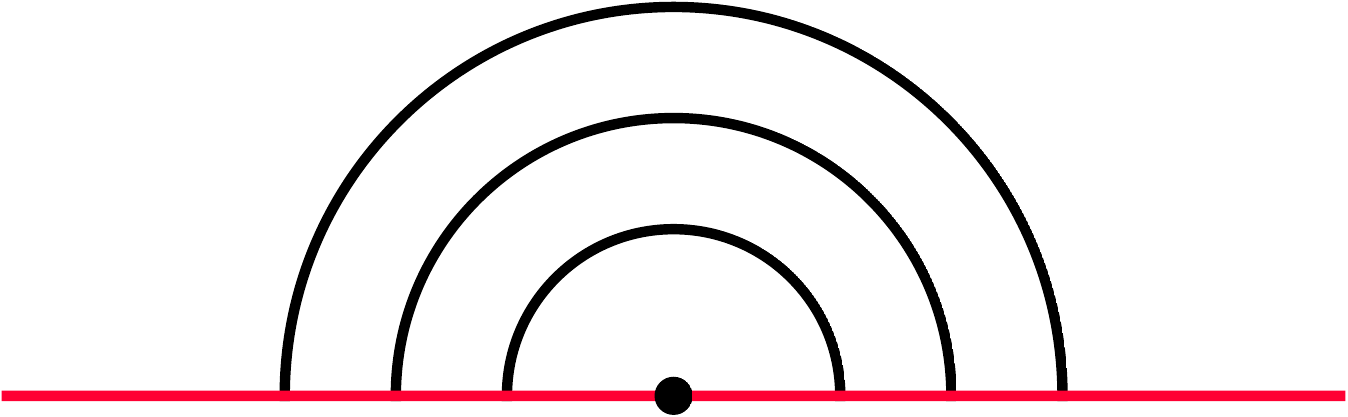}}
    \hspace{5mm}
    \subfigure[$L_1^2$]
   {\includegraphics[width=3.2cm]{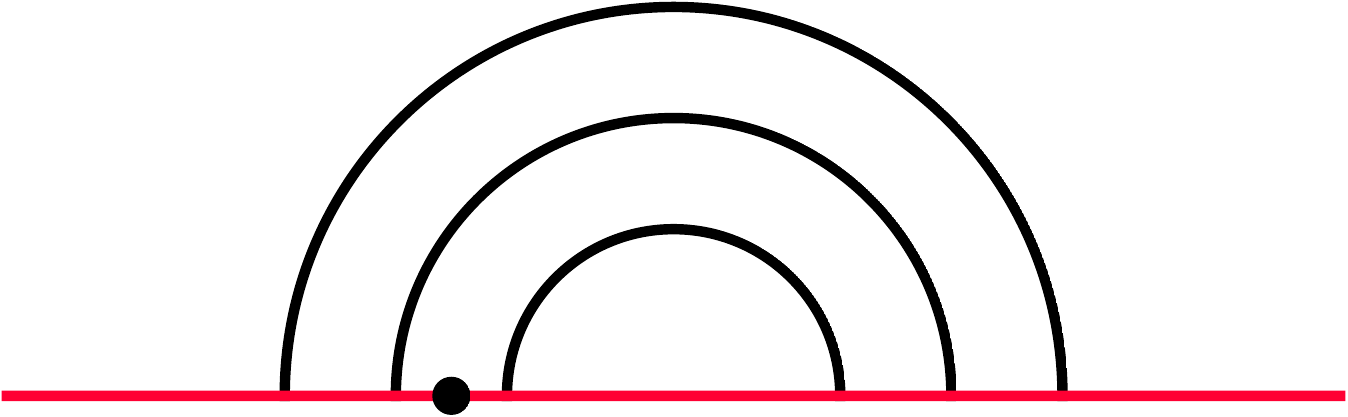}}
    \hspace{5mm}
 \subfigure[$L_1^3$]
   {\includegraphics[width=3.2cm]{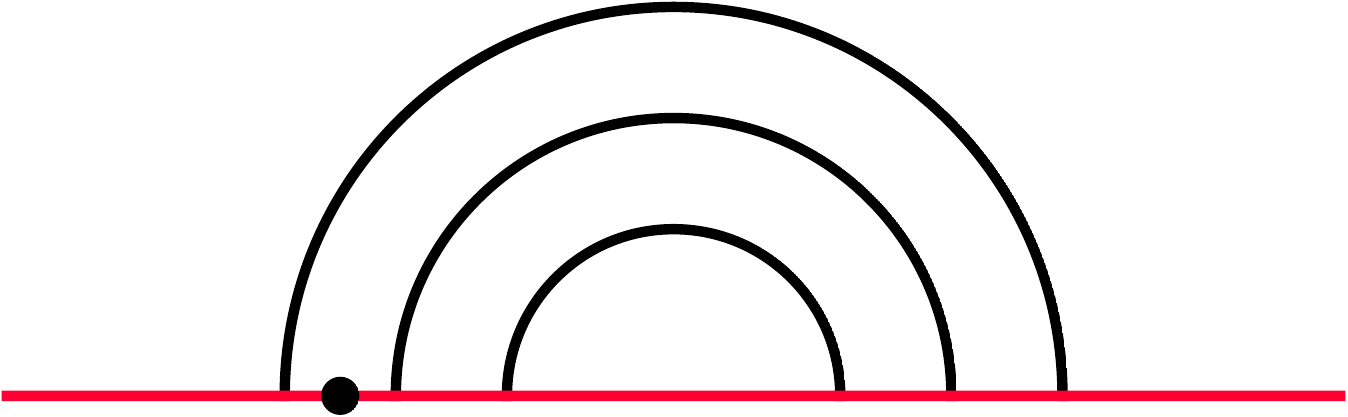}}
    \hspace{5mm}
 \subfigure[$L_1^4$]
   {\includegraphics[width=3.2cm]{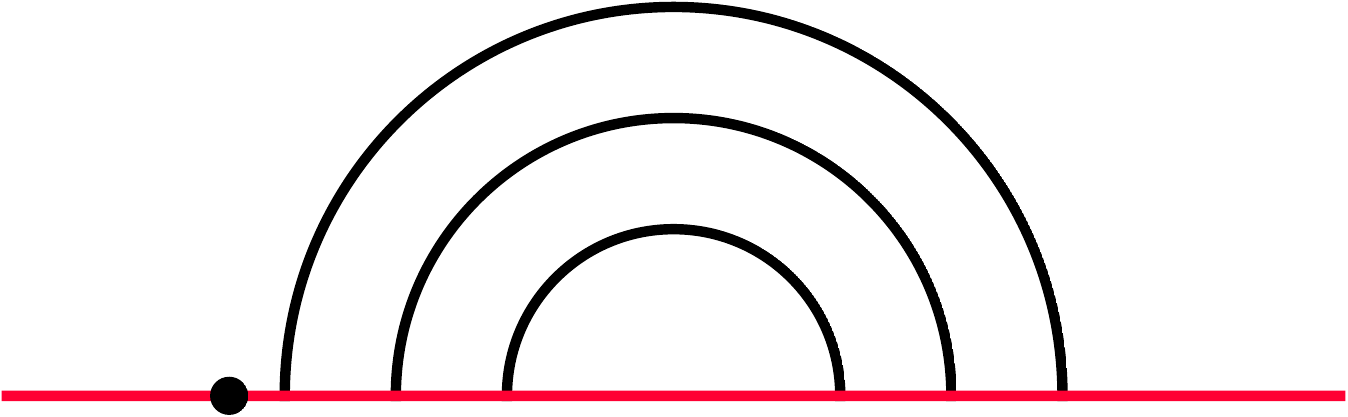}}
 \caption{Fermionic ladder diagrams contributing to the diagram $L_1$.}
  \label{fig:L1graph}
 \end{figure}  
 
 Following the prescriptions of section \ref{sec:sec2} and the legend of 
appendix \ref{sec:appendixB}, the diagram $L_1$ is given by
\begin{equation}\begin{split}\label{L1int}
 L_1^\uparrow=-i\!\left(\frac{2\pi}{\kappa}\right)^3\!\!\!M^2N^2\!\!\!\int_{-L}^L &
 \!\!d\tau_{\mbox{\tiny $\displaystyle1\!\!>\!\! ...\!\!>\!\! 6$}} 
 \biggl[ |\dot{x}_1||\dot{x}_2||\dot{x}_3||\dot{x}_4||\dot{x}_5||\dot{x}_6|(n_1\bar{n}_6)   (\eta_1\gamma^\mu\bar{\eta}_6)   
  \partial^\mu_{x_6}\Delta(x_6,x_1)\\
  &\!\!\!\!\!\!\!\times (n_3\bar{n}_4)   (\eta_3\gamma^\nu\bar{\eta}_4)    \partial^\nu_{x_4}\Delta(x_4,x_3)
  \, (n_5\bar{n}_2)   (\eta_5\gamma^\rho\bar{\eta}_2)   \partial^\rho_{x_2}\Delta(x_2,x_5)\biggl]
\end{split}\end{equation}
Because of the reflection property of the diagrams, we have to compute the integral \eqref{L1int} 
only for the integration regions depicted in Figure \ref{fig:L1graph}, namely 
\begin{align}
&\label{cont1}L_1^1\qquad  (-L<\tau_6<\tau_5<\tau_4<0)\in C_1\cup(0<\tau_3<\tau_2<\tau_1<L)\in C_2\\
&\label{cont2}L_1^2\qquad  (-L<\tau_6<\tau_5<0)\in C_1\cup (0<\tau_4<\tau_3<\tau_2<\tau_1<L)\in C_2\\
&\label{cont3}L_1^3\qquad (-L<\tau_6<0)\in C_1\cup (0<\tau_5<\tau_4<\tau_3<\tau_2<\tau_1<L)\in C_2\\
&\label{cont4}L_1^4\qquad (0<\tau_6<\tau_5<\tau_4<\tau_3<\tau_2<\tau_1<L)\in C_2
\end{align}
then the total result can be 
computed using the following formula
\begin{equation}\label{l1}
L_1^\uparrow=L_1^1+2(L_1^2+L_1^3+L_1^4)\,.
\end{equation}

Consider the diagram $L_1^1$, we substitute $n\bar n$, the bilinears 
$\eta\gamma\bar\eta$ and the propagators $\Delta$ with their values on the 
contour \eqref{cont1} using the the formulas in appendix \ref{sec:appendixA2}, the definitions \eqref{coupl1}, \eqref{coupl2} and 
\eqref{coupl2} and the parametrization \eqref{paramcusp}.
Then the integral \eqref{L1int} becomes
\begin{equation}\label{intL11}
L_1^1\!=\!g(\epsilon)^3 M^2 N^2\! \cos^3\!\frac\theta 2 
\frac{(1\!-2\epsilon)^3}{2^{6\epsilon}}\!\!\!
\int_{0}^1\!\!\!\!  d\tau_{\mbox{\tiny $\displaystyle 1\!\!>\!\! 2\!\!>\!\! 3$}}\!\!
\int_{0}^1\!\!\!\! d\tau_{\mbox{\tiny $\displaystyle 6\!\!>\!\! 5\!\!>\!\! 4$}}
\frac{1}{[(\tau_3+\tau_4)^2(\tau_2+\tau_5)^2(\tau_1+\tau_6)^2]^{1-2\epsilon}}
\end{equation}
where we have rescaled the variables with the dimensional cut-off $L$ as
$\tau\rightarrow -L \tilde{\tau}$ on $C_1$ and $\tau\rightarrow L \tilde{\tau}$ on 
$C_2$. Then for simplicity we have renamed $\tilde\tau$ as $\tau$.
Now the integral \eqref{intL11} can be easily computed and it is given by
\begin{align}
L_1^1=&\frac{g(\epsilon)^3 M^2 N^2\cos^3\!\frac\theta 2}{3\times 2^{4+6\epsilon}\epsilon^3(1+2\epsilon)(1-4\epsilon)(1-6\epsilon)}
\biggl[4 \left(4 \epsilon  \left(2+\epsilon-8 \epsilon ^2\right)-1\right)-64^{\epsilon } (2 \epsilon +1) (1-2 \epsilon 
)^2\nonumber\\
&-3\times 4^{\epsilon +1} \epsilon  (2 \epsilon +1) (6 \epsilon -1)\!-\!12\epsilon(1-6\epsilon)\biggl(\!2^{4 \epsilon +1} (4 \epsilon -1) \, _2F_1(1,4 \epsilon +1;2 (\epsilon +1);-1)\nonumber\\
&+(2 \epsilon +1) \, _2F_1(-2 \epsilon ,4 \epsilon ;4 \epsilon +1;-1)-4 \epsilon  \, _2F_1(1-4 \epsilon ,2 \epsilon +1;2 (\epsilon 
+1);-1)\biggl)\\
&-12 \epsilon  (1-2 \epsilon ) (1-6 \epsilon ) \biggl(16^{\epsilon } \, _3F_2\left(1,1-4 \epsilon ,2-2 \epsilon ;2-4 \epsilon ,2 \epsilon +2;1/2\right)\nonumber\\
&-2 \, _3F_2(1,1-4 \epsilon ,2-2 \epsilon ;2-4 \epsilon ,2 \epsilon 
+2;1)\biggl)\biggl]\nonumber
\end{align}

We repeat the same steps fo the other three diagrams.
The diagram $L_1^2$ has to be computed along the contour \eqref{cont2}, then the 
integral \eqref{L1int} becomes
\begin{equation}\begin{split}
L_1^2\!=&\!g(\epsilon)^3 M^2 N^2\! \cos^2\!\frac\theta 2 
\frac{(1\!-2\epsilon)^3}{2^{6\epsilon}}\!\!\!
\int_{0}^1\!\!\!\!  d\tau_{\mbox{\tiny $\displaystyle 1\!\!>\!\! 2\!\!>\!\! 3\!\!>\!\! 4$}}\!\!
\int_{0}^1\!\!\!\! d\tau_{\mbox{\tiny $\displaystyle 6\!\!>\!\! 5$}}
\frac{1}{[(\tau_3-\tau_4)^2(\tau_2+\tau_5)^2(\tau_1+\tau_6)^2]^{1-2\epsilon}}\\
=&\frac{g(\epsilon)^3 M^2 N^2\cos^2\!\frac\theta 2}{3\times 2^{4+6\epsilon}\epsilon^3(1+2\epsilon)(1-4\epsilon)}
\biggl[2-8\epsilon^2-24 \epsilon ^2 \, _2F_1(1-4 \epsilon ,2 \epsilon +1;2 (\epsilon +1);-1)\\
&-3 (1-4 \epsilon ) ((2 \epsilon +1) \, _2F_1(-2 \epsilon ,4 \epsilon ;4 \epsilon +1;-1)+4^{\epsilon } (4 \epsilon  \, _2F_1(1-2 \epsilon ,2 \epsilon +1;2 (\epsilon 
+1);-1)\\
&-2 \epsilon -1))+6 \epsilon  (1-2 \epsilon) (16^{\epsilon } \, _3F_2(1,1-4 \epsilon ,2-2 \epsilon ;2-4 \epsilon ,2 \epsilon +2;1/2)\\
&-2 \, _3F_2(1,1-4 \epsilon ,2-2 \epsilon ;2-4 \epsilon ,2 \epsilon 
+2;1))\biggl]
\end{split}\end{equation}
The diagram $L_1^3$ has to be computed along the contour \eqref{cont3}, then the 
integral \eqref{L1int} becomes
\begin{equation}\begin{split}
L_1^3\!=&\!g(\epsilon)^3 M^2 N^2\! \cos\frac\theta 2 
\frac{(1\!-2\epsilon)^3}{2^{6\epsilon}}\!\!\!
\int_{0}^1\!\!\!\!  d\tau_{\mbox{\tiny $\displaystyle 1\!\!>\!\! 2\!\!>\!\! 3\!\!>\!\! 4\!\!>\!\! 5$}}\!\!
\int_{0}^1\!\!\!\! d\tau_{\mbox{\tiny $\displaystyle 6$}}
\frac{1}{[(\tau_3-\tau_4)^2(\tau_2-\tau_5)^2(\tau_1+\tau_6)^2]^{1-2\epsilon}}\\
=&\frac{g(\epsilon)^3 M^2 N^2\cos\frac\theta 2}{3\times 2^{4+6\epsilon}}\frac{1-2\epsilon}{\epsilon^3(1-4\epsilon)}
\biggl[1-3\times 4^\epsilon-3 \, _2F_1(-2\epsilon ,4 \epsilon ;1+4\epsilon;-1)\biggl]
\end{split}\end{equation}
The diagram $L_1^4$ has to be computed along the contour \eqref{cont4}, then the 
integral \eqref{L1int} becomes
\begin{equation}\begin{split}
L_1^4\!=&\!g(\epsilon)^3 M^2 N^2
\frac{(1\!-2\epsilon)^3}{2^{6\epsilon}}\!\!\!
\int_{0}^1\!\!\!\!  d\tau_{\mbox{\tiny $\displaystyle 1\!\!>\!\! 2\!\!>\!\! 3\!\!>\!\! 4\!\!>\!\! 5\!\!>\!\! 6$}}
\frac{1}{[(\tau_3-\tau_4)^2(\tau_2-\tau_5)^2(\tau_1-\tau_6)^2]^{1-2\epsilon}}\\
=&-\frac{g(\epsilon)^3 M^2 N^2}{3\times 2^{4+6\epsilon}}\frac{(1-2\epsilon)^2}{\epsilon^3(1-4\epsilon)(1-6\epsilon)}
\end{split}\end{equation}
Summing up the results using the formula \eqref{l1} and expanding the result 
around $\epsilon=0$, the diagram $L_1^\uparrow$ is given by
\footnotesize\begin{equation}\begin{split}
&L_1^\uparrow=g(\epsilon)^3 M^2 N^2
\biggl[
\frac{c_\theta ((c_\theta-2) c_\theta+2)-2}{48 \epsilon 
^3}+\frac{c_\theta (c_\theta (-3 c_\theta (l+1)+6 l+4)-6 l+2)-6}{24 \epsilon 
^2}\\
&+\frac{c_\theta^3 \left(3 l (l+6)+2 \left(\pi ^2-2\right)\right)-2 c_\theta^2 \left(3 l (l+4)+2 \left(\pi ^2-4\right)\right)+2 c_\theta \left(-3 l (l+2)+\pi ^2+4\right)-40}{24 \epsilon 
}\\
&+\!\frac{1}{12} \!\biggl(c_\theta^3\! \left(3 l\! \left(l^2\!+l+4\right)\!-\!2 \pi ^2 (l\!+\!2)\!-\!39 \zeta (3)\!+\!16\right)\!+c_\theta^2\! \left(2 \pi ^2 (l\!+\!1)\!-\!2 l (l (5 l\!+\!12)\!+\!24)\!+\!72 \zeta (3)\!+\!32\right)\\
&+c_\theta \left(-2 l (l (l+3)+12)-33 \zeta (3)+2 \pi ^2+16\right)-128\biggr)+\mathcal{O}(\epsilon)\biggr]
\end{split}\end{equation}\normalsize
where we have introduced the short-hand notation $l=\log 2$ and $c_\theta=\cos\frac{\theta}{2}$.

The remaining fermionic diagrams can be computed in the same way. The 
closed expression and the $\epsilon$-expansion of any diagram can be found in the attached file ``Ladders.nb".
The complete fermionic ladder contribution to the expectation value of the 
cusped Wilson line can be computed with the following formula
\begin{equation}
L_F^\uparrow=L_1^\uparrow+L_2^\uparrow+2L_3^\uparrow+L_5^\uparrow
\end{equation}
and it reads
\footnotesize\begin{align}
&L_F^\uparrow=g(\epsilon)^3 M N\biggl[
\frac{6 (3 c_\theta-4) M^2+(c_\theta ((c_\theta-8) c_\theta+20)-20) M N-2 (c_\theta (2 c_\theta-3)+2) N^2}{48 \epsilon ^3}\nonumber\\
&+\frac{-3 c_\theta^3 (l+1) M N+2 c_\theta^2 N (9 l M+6 l N+5 M+2 N)-2 c_\theta \left(9 l (M+N)^2-4 M N\right)-8 N (3 M+N)}{24 \epsilon ^2}\nonumber\\
&+\frac{1}{72\epsilon}(-2 N^2 \left(9 \left(4 c_\theta^2 (l-1)+3 c_\theta l^2+8\right)+\pi ^2 (c_\theta (7 c_\theta-6)-2)\right)+3 M N (c_\theta^3 \left(3 l (l+6)+2 \left(\pi ^2-2\right)\right)\nonumber\\
&+c_\theta^2 \left(6 (l-8) l-6 \pi ^2+40\right)-8 c_\theta \left(9 l^2+3 l-4\right)+8 \left(\pi ^2-14\right))-6 M^2 \left(27 c_\theta l^2+\pi ^2 (5 c_\theta-8)\right))\nonumber\\
&+\!\frac{1}{36} (3 c_\theta^3 M N \left(3 \left(l^2\!+\!l\!+\!4\right) l\!-\!2 \pi ^2 l\!-\!39 \zeta (3)\!-\!4 \pi ^2\!+\!16\right)\!-c_\theta^2 N (3 M (2 l \left(7 l^2\!+30 l+3 \pi ^2+48\right)\!-\!75 \zeta (3)\nonumber\\
&-80)+2 N \left(3 l \left(6 l (l+3)+\pi ^2+36\right)+2 \left(-51 \zeta (3)+\pi ^2-54\right)\right))+6 c_\theta (l^3 \left(-\left(27 M^2+26 M N+3 N^2\right)\right)\nonumber\\
&-12 l^2 M N+l \!\left(\pi ^2 \!\left(9 M^2\!+\!10 M N\!+\!2 N^2\right)\!-\!24 M N\right)\!+\!32 M N\!+\!6 \zeta (3) \left(12 M^2\!+\!3 M N\!-\!5 N^2\right))\nonumber\\
&+8 N \left(3 \left(\pi ^2\!-\!34\right) \!M\!+\!\left(\pi ^2\!-\!54\right)\! N\right)-48 \zeta (3) \left(15 M^2+9 M N+N^2\right))+\mathcal{O}(\epsilon)\biggr]
\end{align}\normalsize

\paragraph{Bosonic ladder diagrams}$\\$

The second class of diagrams can be computed performing the planar Wick contractions 
of the mixed bosonic-fermionic monomial \eqref{t2}. In order to do it using the package \texttt{WiLE} we have to set the
initial data to
\begin{equation}
(n_f=2,n_g=0,n_s=2)\qquad\text{and}\qquad V_i=0\quad \text{with}\quad 
i=1,2,...,7\,.
\end{equation} 
As the previoue case, we set the options \texttt{Supermatrix Sector} and \texttt{Planar} 
on \texttt{UL} and \texttt{on} respectively and the remaining options on \texttt{off}. 
Given all the initial data the \texttt{WiLE} output is the following
\vspace{-.4cm}
\begin{center}
\noindent\hspace{.5cm}\includegraphics[trim=1.6cm 0 0 0, clip=true, scale=.31]{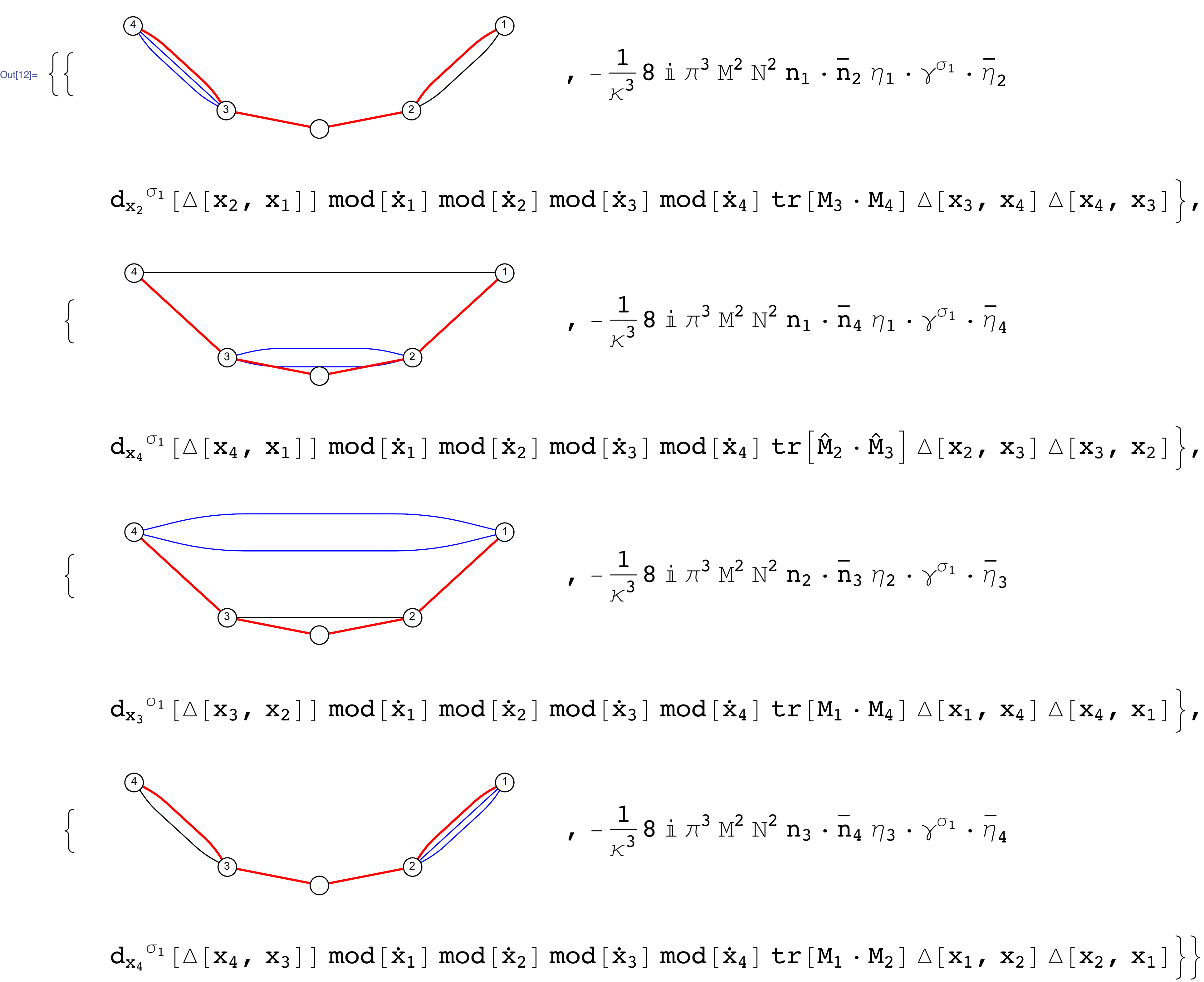}
\end{center}
\vspace{-.4cm} 
In order to simplify the following discussion we name the diagrams above $L^\uparrow_i$ $i=6,...,9$ 
where the index $i$ corresponds to the raw of the \texttt{WiLE} output.
For this class of diagrams, we have to split the integration on the Wilson line in five 
subregions. Using the reflection property of the diagrams we have 
\begin{equation}
L^\uparrow_9=L^\uparrow_6\,.
\end{equation}
We will compute in detail the diagram $L_6$ as an example.

\begin{figure}[!t]
 \centering
  \subfigure[$L_6^1$]
   {\includegraphics[width=3.2cm]{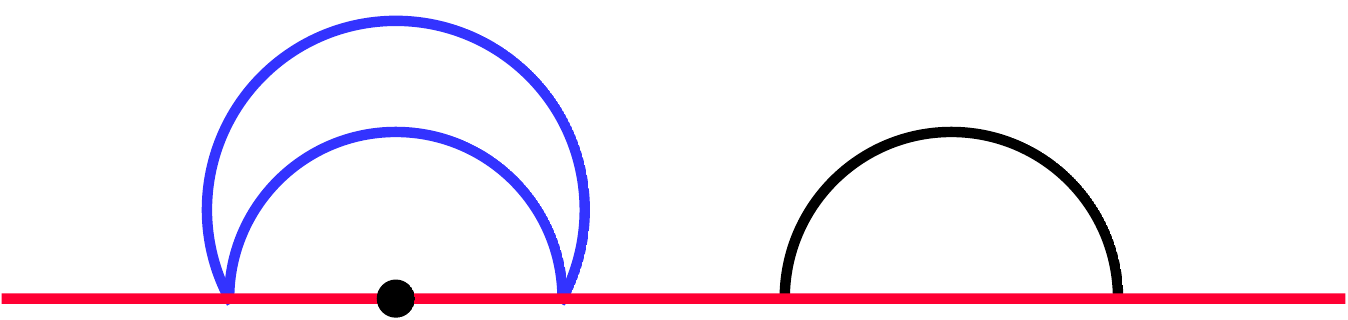}}
    \hspace{5mm}
    \subfigure[$L_6^2$]
   {\includegraphics[width=3.2cm]{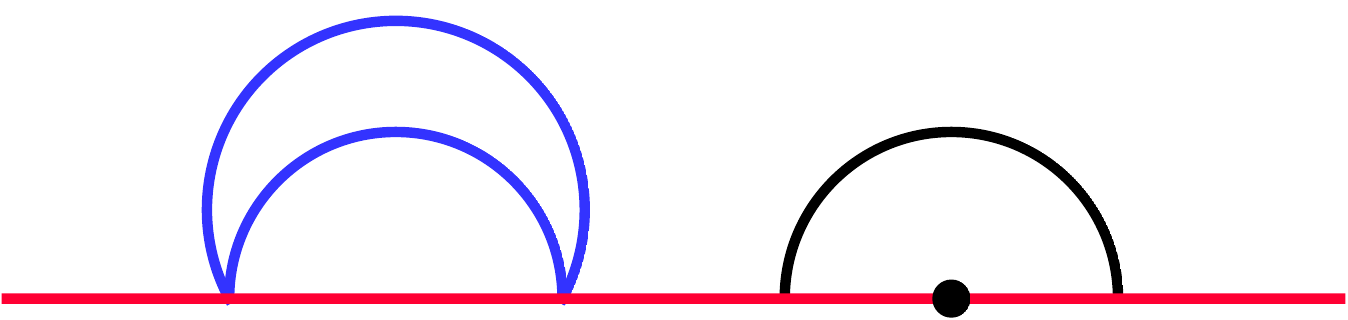}}
    \hspace{5mm}
 \subfigure[$L_6^3$]
   {\includegraphics[width=3.2cm]{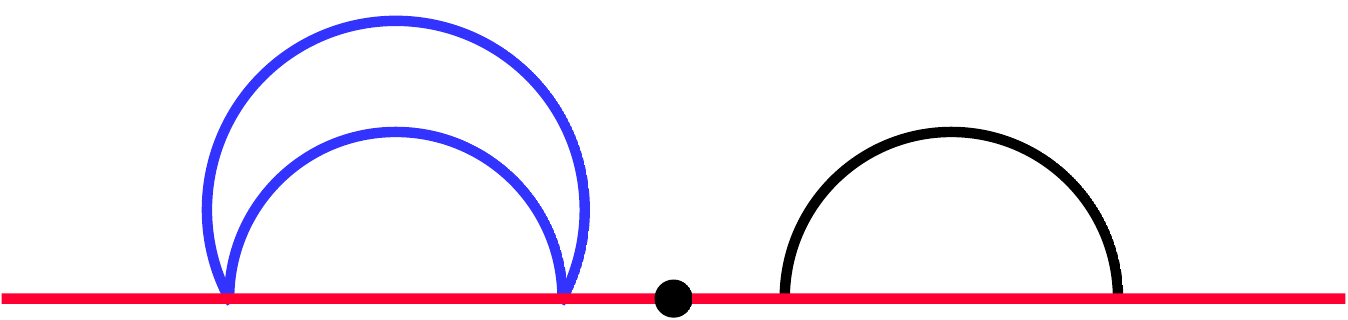}}
    \hspace{5mm}
 \subfigure[$L_6^4$]
   {\includegraphics[width=3.2cm]{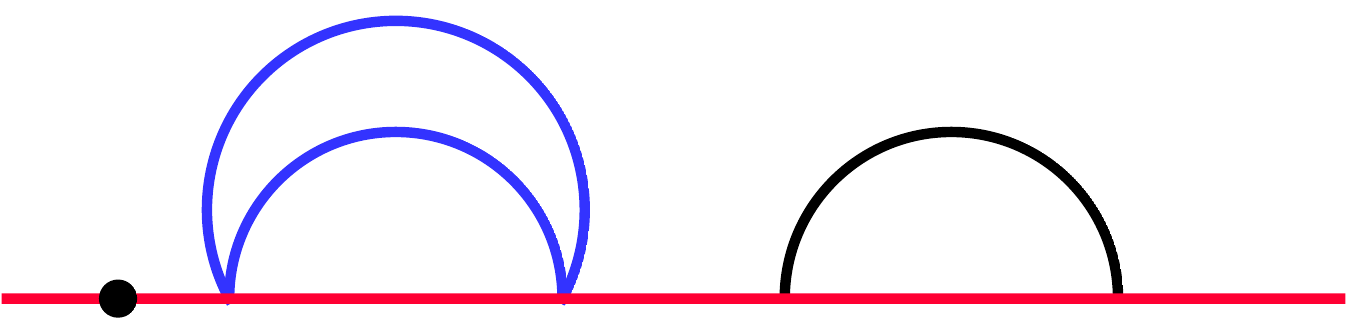}}
 \caption{Bosonic ladder diagrams contributing to the diagram $L_6$.}
  \label{fig:L6graph}
 \end{figure}

 Following the prescriptions of section \ref{sec:sec2} and the legend of 
appendix \ref{sec:appendixB}, the diagram $L_6$ is given by
\small\begin{equation}\begin{split}\label{L6int}
 L_6^\uparrow=-i\!\left(\frac{2\pi}{\kappa}\right)^3\!\!\!M^2N^2\!\!\!\int_{-L}^L &
 \!\!d\tau_{\mbox{\tiny $\displaystyle1\!\!>\!\! ...\!\!>\!\! 4$}} 
 \biggl[ |\dot{x}_1||\dot{x}_2||\dot{x}_3||\dot{x}_4|\text{tr}(\mathcal{M}_3\mathcal{M}_4)(n_1\bar{n}_2)   (\eta_1\gamma^\mu\bar{\eta}_2)   
  \partial^\mu_{x_2}\Delta(x_2,x_1)\Delta^2(x_3,x_4)\biggl]
\end{split}\end{equation}\normalsize
Because of the reflection property of the diagrams, we have to compute the integral \eqref{L6int} 
only for the integration regions depicted in Figure \ref{fig:L6graph}, namely 
\begin{align}
&\label{cont61}L_6^1\qquad  (-L<\tau_4<0)\in C_1\cup(0<\tau_3<\tau_2<\tau_1<L)\in C_2\\
&\label{cont62}L_6^2\qquad  (-L<\tau_4<\tau_3<\tau_2<0)\in C_1\cup (0<\tau_1<L)\in C_2\\
&\label{cont63}L_6^3\qquad (-L<\tau_4<\tau_3<0)\in C_1\cup (0<\tau_2<\tau_1<L)\in C_2\\
&\label{cont64}L_6^4\qquad (0<\tau_4<\tau_3<\tau_2<\tau_1<L)\in C_2
\end{align}
then the total result can be 
computed using the following formula
\begin{equation}\label{l6}
L_6^\uparrow=L_6^1+L_6^2+L_6^3+2L_6^4\,.
\end{equation}
Following the same steps of the fermionic case, the integral \eqref{L6int} along the contour \eqref{cont61} is given by
\begin{equation}\begin{split}\label{intL61}
L_6^1\!&=\!g(\epsilon)^3 M^2 N^2\! \cos^2\!\frac\theta 2 
\frac{(1\!-2\epsilon)}{2^{6\epsilon}}\!\!\!
\int_{0}^1\!\!\!\!  d\tau_{\mbox{\tiny $\displaystyle 1\!\!>\!\! 2\!\!>\!\! 3$}}\!\!
\int_{0}^1\!\!\!\! d\tau_{\mbox{\tiny $\displaystyle 4$}}
\frac{1}{[(\tau_1-\tau_2)^2]^{1-\epsilon}}\frac{1}{[(\tau_3+\tau_4)^2]^{1-2\epsilon}}\\
&=\!-\frac{g(\epsilon)^3 M^2 N^2\cos^2\!\frac\theta 2}{2^{3+6\epsilon}\epsilon^2(1+2\epsilon)(1-4\epsilon)}
\biggl[1\!+\!\frac{\Gamma(1\!+\!2\epsilon)\Gamma(1\!+\!4\epsilon)}{\Gamma(1\!+\!6\epsilon)}-\, _2F_1(1,-4 \epsilon ;1+2 
\epsilon;-1))\biggr]
\end{split}\end{equation}
The diagram $L_6^2$ has to be computed along the contour \eqref{cont62} and it reads
\begin{equation}\begin{split}\label{intL62}
L_6^2\!&=\!g(\epsilon)^3 M^2 N^2\! \cos\!\frac\theta 2 
\frac{(1\!-2\epsilon)}{2^{6\epsilon}}\!\!\!
\int_{0}^1\!\!\!\!  d\tau_{\mbox{\tiny $\displaystyle 1$}}\!\!
\int_{0}^1\!\!\!\! d\tau_{\mbox{\tiny $\displaystyle 4\!\!>\!\! 3\!\!>\!\! 2$}}
\frac{1}{[(\tau_1+\tau_2)^2]^{1-\epsilon}}\frac{1}{[(\tau_4-\tau_3)^2]^{1-2\epsilon}}\\
&=\!-\frac{g(\epsilon)^3 M^2 N^2\cos\!\frac\theta 2}{2^{3+6\epsilon}\epsilon^2(1-4\epsilon)}
\biggl[1\!+\!\frac{\Gamma(1\!+\!2\epsilon)\Gamma(1\!+\!4\epsilon)}{\Gamma(1\!+\!6\epsilon)}-\, _2F_1(1,-2 \epsilon ;1+4 
\epsilon;-1))\biggr]
\end{split}\end{equation}
The diagram $L_6^3$ has to be computed along the contour \eqref{cont63} and it reads
\begin{equation}\begin{split}\label{intL63}
L_6^3\!&=\!g(\epsilon)^3 M^2 N^2
\frac{(1\!-2\epsilon)}{2^{6\epsilon}}\!\!\!
\int_{0}^1\!\!\!\!  d\tau_{\mbox{\tiny $\displaystyle 1\!\!>\!\! 2$}}\!\!
\int_{0}^1\!\!\!\! d\tau_{\mbox{\tiny $\displaystyle 4\!\!>\!\! 3$}}
\frac{1}{[(\tau_1-\tau_2)^2]^{1-\epsilon}}\frac{1}{[(\tau_4-\tau_3)^2]^{1-2\epsilon}}\\
&=\frac{g(\epsilon)^3 M^2 N^2}{2^{3+6\epsilon}\epsilon^2(1-4\epsilon)}
\end{split}\end{equation}
Finally the diagram $L_6^4$ has to be computed along the contour \eqref{cont64} and it is given by
\begin{equation}\begin{split}\label{intL64}
L_6^4\!&=\!g(\epsilon)^3 M^2 N^2
\frac{(1\!-2\epsilon)}{2^{6\epsilon}}\!\!\!
\int_{0}^1\!\!\!\!  d\tau_{\mbox{\tiny $\displaystyle 1\!\!>\!\! 2\!\!>\!\! 3\!\!>\!\! 4$}}
\frac{1}{[(\tau_1-\tau_2)^2]^{1-\epsilon}}\frac{1}{[(\tau_3-\tau_4)^2]^{1-2\epsilon}}\\
&=-\frac{g(\epsilon)^3 M^2 N^2}{2^{6\epsilon}}\frac{\Gamma(2\epsilon)\Gamma(4\epsilon-1)}{\Gamma(1+6\epsilon)}
\end{split}\end{equation}
Summing up the results using the formula \eqref{l6} and expanding the result 
around $\epsilon=0$, the diagram $L_6^\uparrow$ is given by
\begin{equation}\begin{split}
L_6^\uparrow=g(\epsilon)^3& M^2 N^2\biggl[\frac{3-c_\theta (c_\theta+1)}{8 \epsilon ^2}+\frac{c_\theta (2 c_\theta (l-1)+l-2)+6}{4 \epsilon }\\
&+\frac{1}{12} \left(\pi ^2 \left(c_\theta^2+c_\theta-4\right)+3 c_\theta (2 c_\theta (l+2) (3 l-2)+l (3 l+4)-8)+72\right)\biggr]
\end{split}\end{equation}
The remaining bosonic diagrams can be computed in the same way. The 
closed expression and the $\epsilon$-expansion of any diagram can be found in the attached file ``Ladders.nb".
The complete bosonic ladder contribution to the expectation value of the 
cusped Wilson line can be computed with the following formula
\begin{equation}
L_B^\uparrow=2L_6^\uparrow+L_7^\uparrow+L_8^\uparrow
\end{equation}
and it reads
\footnotesize\begin{equation}\begin{split}
L_B^\uparrow=&g(\epsilon)^3 M^2 N^2\biggl[\frac{(c_\theta-4) c_\theta (3 c_\theta+2)+24}{24 \epsilon ^2}+\frac{3 c_\theta (c_\theta (8-3 c_\theta)+4) l+2 (c_\theta ((c_\theta-10) c_\theta-8)+28)}{12 \epsilon }\\
&-\frac{1}{12} \left(c_\theta^3\left(3 l (l+4)-4 \left(\pi ^2-4\right)\right)-c_\theta^2 (12 l (5 l+8)-80)-8 c_\theta (3 l (l+2)-8)+8 \left(\pi ^2-34\right)\right)\biggr]
\end{split}\end{equation}\normalsize

\paragraph{Summing up fermionic and bosonic ladders}

The total contribution of the ladder diagrams to the three-loop expectation value of the upper-left block of the Wilson loop can be computed using the following formula
\begin{equation}
\langle\mathcal{W}^\uparrow[C]\rangle^{(3)}_{ladder}=L_F^\uparrow+L_B^\uparrow\,.
\end{equation}
Recalling the prescription give in \eqref{vevpm} and the relation \eqref{symvev}, we are finally able to compute the expectation value of the fermionic operators $\mathcal{W}_\pm$. The traced operator $\mathcal{W}_+$ is given by
\begingroup\makeatletter\def\f@size{9}\check@mathfonts
\begin{equation}\begin{split}\label{Wtrfin}
&\langle\mathcal{W}_+[C]\rangle^{(3)}_{ladder}=g(\epsilon)^3\frac{MN}{N+M}\biggl[\frac{-2 ((c_\theta-6) c_\theta+7) (M^2+N^2)+(c_\theta ((c_\theta-8) c_\theta+20)-20) M N}{24 \epsilon ^3}\\
&+\frac{2 (M^2\!+N^2) \left(c_\theta^2+3 (c_\theta-3) c_\theta l-2\right)-3 c_\theta ((c_\theta-6) c_\theta+\!12) l M N}{12 \epsilon ^2}-\frac{1}{36 \epsilon }((M^2+N^2) (36 (c_\theta^2 (l-1)\\
&+3 c_\theta l^2+2)+\pi ^2 (c_\theta (7 c_\theta+9)-26))-3 M N \left(3 (c_\theta-4) c_\theta (c_\theta+6) l^2+2 \pi ^2 (c_\theta+1) (c_\theta-2)^2\right))\\
&-\frac{1}{18} ((M^2+N^2)(18 c_\theta \left(c_\theta (l (l (l+3)+6)-6)+5 l^3\right)-\pi ^2 c_\theta (33 l-c_\theta (3 l+2))\\
&-6 (c_\theta (17 c_\theta+21)-64) \zeta (3)-4 \pi ^2+216)+3MN( c_\theta l((c_\theta (14-3 c_\theta)+52) l^2\\
&+2 \pi ^2 (c_\theta-2) (c_\theta+5))+3 (c_\theta (c_\theta (13 c_\theta-25)-12)+48)\zeta (3)))+\mathcal{O}(\epsilon)\biggr]\,,
\end{split}\end{equation}\endgroup

and the supertraced operator $\mathcal{W}_-$ is given by
\begingroup\makeatletter\def\f@size{9}\check@mathfonts
\begin{equation}\begin{split}\label{Wstrfin}
&\langle\mathcal{W}_-[C]\rangle^{(3)}_{ladder}=g(\epsilon)^3MN(M+N) \biggl[-\frac{(c_\theta (c_\theta+3)-5)}{12 \epsilon ^3}+\frac{\left(c_\theta^2 (3 l+1)-2\right)}{6 \epsilon ^2}\\
&-\frac{\left(18 \left(2 c_\theta^2 (l-\!1)-3 c_\theta l^2\!+4\right)+\pi ^2 (7 (c_\theta\!-3) c_\theta+22)\right)}{36 \epsilon }-\frac{1}{18} (18 c_\theta \left(c_\theta (l (l (l+3)+6)-6)-4 l^3\right)\\
&+\pi ^2 c_\theta (3 (c_\theta+7) l+2 c_\theta)-(102 (c_\theta-3) c_\theta+336) \zeta (3)-4 \pi ^2+216)+\mathcal{O}(\epsilon)\biggr]\,.
\end{split}\end{equation}\endgroup
The divergent part of \eqref{Wtrfin}, in the $M=N$ case, is in agreement with the same computation done in ABJM using the heavy quark effective theory formalism (HQET) in \cite{Bianchi:2017svd}.

Recently, the ladder contribution to the expectation value of the cusped Wilson loops 
$\mathcal{W}_\pm$ was computed in \cite{Bonini:2016fnc} using the Bethe-Salpeter equation in the following scaling limit
\begin{equation}\label{scaling}
i\theta>>0\,,\quad\lambda_{1,2}<<1\qquad \text{with}\;\;\hat{\lambda}_{1,2}=\lambda_{1,2}\cos\frac{\theta}{2}\quad\text{fixed}\,,
\end{equation}
where $\lambda_{1,2}$ are the 't Hooft coupling defined in \eqref{thooft}. In this limit only the leading divergence of the ladders contributes and the perturbative series can be reorganized in powers of $\hat\lambda$. This allows to compute exactly $\mathcal{W}_\pm$ for the cusped contour. In the $\varphi=0$ case the solutions are the following
\begin{equation}\begin{split}\label{surprises}
\langle\mathcal{W}_+[C]\rangle=&\frac{(\sqrt{M}+\sqrt{N})^2}{2(N+M)}e^{\frac{1}{2\epsilon}\frac{\sqrt{N M}c_\theta}{\kappa}}+\frac{(\sqrt{M}-\sqrt{N})^2}{2(N+M)}e^{-\frac{1}{2\epsilon}\frac{\sqrt{N M}c_\theta}{\kappa}}\,,\\
\langle\mathcal{W}_-[C]\rangle=&\frac12e^{\frac{1}{2\epsilon}\frac{\sqrt{N M}c_\theta}{\kappa}}+\frac12e^{-\frac{1}{2\epsilon}\frac{\sqrt{N M}c_\theta}{\kappa}}\,.
\end{split}\end{equation}  
Considering the scaling limit \eqref{scaling}, the leading divergences of \eqref{Wtrfin} and \eqref{Wstrfin} are
\begin{equation}
\langle\mathcal{W}_+[C]\rangle^{(3)}_{ladder}\overset{s.l.}=\frac{M^2N^2c_\theta^3}{24(M+N)\epsilon^3\kappa^3}\qquad\text{and}\quad\langle\mathcal{W}_-[C]\rangle^{(3)}_{ladder}\overset{s.l.}=0\,,
\end{equation}
that are in agreement with the three-loop expansion of \eqref{surprises}.

\section*{Acknowledgments}
It is a pleasure to thank Lorenzo Bianchi, Luca Griguolo, Marco Meineri, Domenico Seminara and Edoardo Vescovi for very useful discussions and the critical reading of the draft. The work of MP is supported by  "Della Riccia Foundation" grant.

\appendix

\section{Notation and conventions}\label{sec:appendixA}

We work in Euclidean space in three dimensions with coordinates $x^\mu = \{x^1, x^2, x^3\}$. We choose a set of Dirac gamma matrices satisfying the three dimensional Clifford algebra as
\begin{equation}\label{gammam}
(\gamma^\mu)_\alpha^{\; \, \beta} = \{ -\sigma^3, \sigma^1, \sigma^2 \}
\end{equation}
where $\sigma^i$ are the Pauli matrices.
In our notation, the product of matrices is given by
\begin{equation}
\label{prod}
(\gamma^\mu \gamma^\nu)_\alpha^{\; \, \beta} \equiv (\gamma^\mu)_\alpha^{\; \, \gamma} (\gamma^\nu)_\gamma^{\; \, \beta}
\end{equation}
The gamma matrices \eqref{gammam} satisfy the following relations
\begin{equation}\begin{split}\label{gammaprod}
\gamma^\mu \gamma^\nu &= \delta^{\mu \nu} \mathbb{1} - i \epsilon^{\mu\nu\rho} \gamma^\rho\;\quad\qquad\qquad\qquad\qquad\qquad\text{tr}(\gamma^\mu \gamma^\nu) = 2 \delta^{\mu\nu}\\
\gamma^\mu \gamma^\nu \gamma^\rho &= \delta^{\mu\nu} \gamma^\rho - \delta^{\mu\rho} \gamma^\nu+  \delta^{\nu\rho} \gamma^\mu  - i \epsilon^{\mu\nu\rho} \mathbb{1}\quad\qquad\text{tr}(\gamma^\mu \gamma^\nu \gamma^\rho) = -2i \epsilon^{\mu\nu\rho}\,.
\end{split}\end{equation}
Spinor indices are lowered and raised with the usual $\epsilon$-tensor
\begin{equation}
\psi^\alpha = \epsilon^{\alpha\beta} \psi_\beta \quad , \quad \psi_\alpha = \epsilon_{\alpha\beta}  \psi^\beta    
\end{equation}
where $\epsilon^{12} = - \epsilon_{12} = 1$.
Moreover, under complex conjugation the Dirac matrices transform as follows:
\begin{equation}
[(\gamma^\mu)_\alpha^{\; \, \beta}]^* = (\gamma^\mu)^\beta_{\; \, \alpha} \equiv \epsilon^{\beta\gamma} (\gamma^\mu)_\gamma^{\; \, \delta}  \epsilon_{\alpha\delta}
\end{equation}

During the evaluation of fermionic contributions to the Wilson loop vev, we often encounter bilinears constructed with the spinors $\eta$ and $\bar \eta$  defined by the two relations in \eqref{condizioneeta}. The simplest bilinear contains one Dirac gamma matrix and it can be expressed as a function of the position as follows
\begin{equation}
\label{rg6}
\begin{split}
(\eta_{2}\gamma^{\mu}\bar\eta_{1})
=-\frac{2}{(\eta_{1}\bar\eta_{2})}\left[\frac{\dot{x_{1}}^{\mu}}{|\dot x_{1}|}+\frac{\dot{x_{2}}^{\mu}}{|\dot{x}_{2}|}-i\frac{\dot{x_{2}}^{\lambda}}{|\dot{x}_{2}|}
\frac{\dot{x_{1}}^{\nu}}{|\dot x_{1}|}\epsilon_{\lambda\nu}^{\ \  \ \mu}
\right].
\end{split}
\end{equation}
where the superscripts $1$ and $2$  denote two different points on the contour.
Bilinears with a higher number of contracted gamma matrices can be rewritten in terms of the bilinear \eqref{rg6} and the spinors contraction $\eta\bar\eta$ using the relations \eqref{gammaprod} and their generalization.

The $U(N)$ gauge group generators $T^A=(T^0,T^a)$, where $T^0=\tfrac{\mathbb{1}}{\sqrt{N}}$ and $a=1,...,N^2-1$, are an orthonormal set of traceless $N\times N$ hermitian matrices with the following normalization
\begin{equation}
\text{Tr}_{\mathbf{N}}(T^AT^B)=\delta^{AB}
\end{equation}
similarly for the set of $M\times M$ generators of $U(M)$.
The \texttt{WiLE} package uses the double line notation in which the fields carry two indices in the fundamental representation of the two gauge groups. In general the lowercase roman indices $i, j,...$ are in the fundamental representation of $U(N)$ and the hatted lowercase roman indices $\hat{i}, \hat{j},...$ are in the fundamental representation of $U(M)$.

\section{The ABJM theory action}\label{sec:appendixA2}
 
 \begin{figure}[!h]
\centering
	\includegraphics[width=10cm]{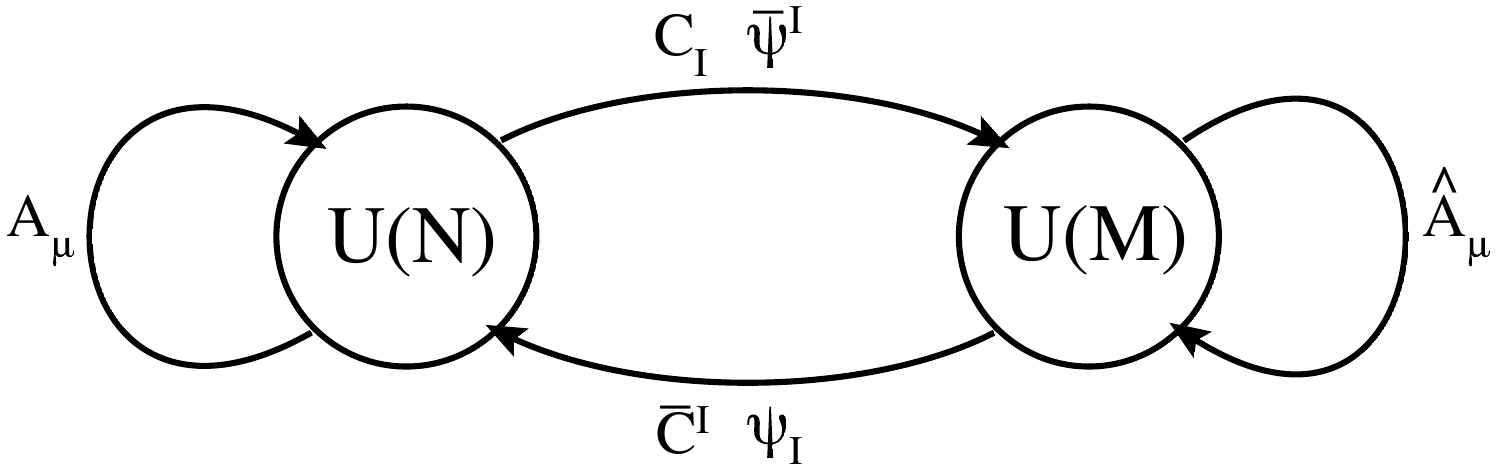}
	\caption{Quiver diagram for ABJ(M) theory.}
		\label{fig:quiver}
\end{figure}

The $\mathcal{N}=6$ super Chern-Simons-matter theory, also known as ABJ(M) in three dimensions, is a superconformal field theory with gauge group $U_{\kappa}(N)\times U_{-\kappa}(M)$ and global symmetry given by the orthosymplectic supergroup $OSp(6|4)$ \cite{Aharony:2008ug,Bandres:2008ry}. The bosonic part of the supergroup involves the R-symmetry group $SO(6)\sim~SU(4)$ and the three-dimensional 
conformal group $Sp(4)\sim SO(2,3)$. The fermionic part generates the $\mathcal{N}=6$ supersymmetries.

The field content of the ABJ(M) theory can be schematically represented by the quiver 
in Figure \ref{fig:quiver}. 
The gauge fields ${(A_\mu)_i}^j$ and 
${(\hat{A}_\mu)_{\hat{i}}}^{\hat{j}}$, belonging respectively to
the adjoint of $U_{\kappa}(N)$ and $U_{-\kappa}(M)$, are represented by arrows forming a loop with their gauge group sector. 
The matter sector instead contains a couple of complex scalars
${(C_I)_i}^{\hat{j}}$ and ${(\bar{C}^I)_{\hat{i}}}^j$ as well as the 
fermions ${(\psi_{I})_{\hat{i}}}^j$ and ${(\bar{\psi}^I)_i}^{\hat{j}}$.
The matter fields are represented by arrows connecting the two different side of the gauge group $U_{\kappa}(N)\times U_{-\kappa}(M)$, indeed
the fields $(C, \bar\psi)$ transform in the $(\mathbf{N},\bar{\mathbf{M}})$ representation  while the pair $(\bar{C},\psi)$ in the $(\bar{\mathbf{N}},\mathbf{M})$. The additional capital index $I=1,...,4$ belongs to the R-symmetry group 
$SU(4)$. We have also to introduce the covariant
gauge fixing function $\partial_\mu A^\mu$ for both gauge fields and two sets of 
ghosts $(\bar{c},c)$ and $(\bar{\hat{c}},\hat{c})$, in order to quantize the theory at the perturbative level. 
Then the ABJ(M) action can be written as the sum of the following terms
\begin{equation}\label{abjmactiontot}
S_{\text{ABJ(M)}}=S_{\text{CS}}+S_{\text{gf}}+S_{\text{matter}}+S^{F}_{\text{pot}}+S^{B}_{\text{pot}}
\end{equation}
namely the Chern-Simons action, the gauge-fixing action, the matter term, 
a fermionic and a bosonic potential. We work with the following 
Euclidian space action
\small\begin{align}
\label{Scs}S_{\text{CS}}=&-i\frac{\kappa}{4\pi}\int d^3x \epsilon^{\mu\nu\rho}\left[\text{Tr}(A_\mu\partial_\nu A_\rho+\tfrac 23 i A_\mu A_\nu A_\rho)-
\text{Tr}(\hat{A}_\mu\partial_\nu \hat{A}_\rho+\tfrac 23 i \hat{A}_\mu \hat{A}_\nu 
\hat{A}_\rho)\right],\\
\label{Sgf}S_{\text{gf}}=&\frac{\kappa}{4\pi}\int d^3x 
\left[\tfrac 1\xi \text{Tr}\left(\partial_\mu A^\mu\right)^2+\text{Tr}\left(\partial_\mu \bar{c} D^\mu c\right)
-\tfrac 1\xi \text{Tr}(\partial_\mu \hat{A}^\mu)^2-\text{Tr}\left(\partial_\mu \bar{\hat{c}} D^\mu 
\hat{c}\right)\right],\\
\label{Smatter}S_{\text{matter}}=&\int d^3x \left[\text{Tr}\left(D_\mu C_I D^\mu 
\bar{C}^I\right)+i \text{Tr}\left(\bar{\psi}^I\slashed{D}\psi_I\right)\right],
\end{align}\normalsize
and the following potentials
\begin{align}
\label{SF}S^{F}_{\text{pot}}=&-\frac{2\pi i}{\kappa}\int d^3x
\left[ \text{Tr}(\bar{C}^I C_I \psi_J \bar{\psi}^J)-\text{Tr}( C_I 
\bar{C}^I\bar{\psi}^J\psi_J)+2\text{Tr}( C_I \bar{C}^J\bar{\psi}^I\psi_J)\right.\nonumber\\
&\qquad\left. -2\text{Tr}(\bar{C}^I C_J \psi_I 
\bar{\psi}^J)-\epsilon_{IJKL}\text{Tr}(\bar{C}^I\bar{\psi}^J\bar{C}^K\bar{\psi}^L)+\epsilon^{IJKL}\text{Tr}(C_I\psi_J C_K 
\psi_L)\right],\\
\label{SB}S^{B}_{\text{pot}}=&-\frac{4\pi^2}{3\kappa^2}\int d^3x
\left[ \text{Tr}(C_I\bar{C}^I C_J\bar{C}^J C_K\bar{C}^K)+\text{Tr}(\bar{C}^I C_I \bar{C}^J C_J \bar{C}^K C_K)\right.\nonumber \\
&\qquad\left. +4\text{Tr}(C_I\bar{C}^J C_K\bar{C}^I C_J\bar{C}^K)-6 \text{Tr}(C_I\bar{C}^J C_J\bar{C}^I 
C_K\bar{C}^K)\right],
\end{align}
where $\epsilon^{1234}=\epsilon_{1234}=1$ and $\kappa$ is the integer Chern-Simons level.
The matter covariant derivatives are defined as
\begin{equation}\begin{split}
D_\mu C_I=&\partial_\mu C_I+i(A_\mu C_I-C_I\hat{A}_\mu),\qquad
D_\mu \bar{C}^I=\partial_\mu\bar{C}^I-i(\bar{C}^I A_\mu - 
\hat{A}_\mu\bar{C}^I),\\
D_\mu \psi_I=&\partial_\mu \psi_I+i(\hat{A}_\mu \psi_I-\psi_I A_\mu),\qquad\;
D_\mu \bar{\psi}^I=\partial_\mu\bar{\psi}^I-i(\bar{\psi}^I \hat{A}_\mu-A_\mu \bar{\psi}^I).
\end{split}\end{equation}
The interaction vertices $V_i$ can be read directly from the action \eqref{abjmactiontot} as mentioned in the section \ref{sec:manual}.

The action given by \eqref{abjmactiontot} is invariant under the following supersymmetry transformations
\small\begin{align}\label{susytransfABJ}
\delta A_\mu&=\frac{4\pi i}{k} \bar \theta^{IJ\alpha} (\gamma_\mu)_\alpha^{\ \beta}
\left(C_I \psi_{J\beta}+\frac{1}{2}\epsilon_{IJKL}\bar \psi^K_\beta \bar C^L\right)\,,
\cr
\delta  \hat{A}_\mu&=\frac{4\pi i}{k} \bar \theta^{IJ\alpha} (\gamma_\mu)_\alpha^{\ \beta}
\left(\psi_{J\beta} C_I+\frac{1}{2}\epsilon_{IJKL}\bar C^L\bar \psi^K_\beta\right)\,,
\cr
\delta C_K&= \bar \theta^{IJ\alpha} \epsilon_{IJKL} \bar \psi^L_\alpha\,,
\\
\delta \bar C^K&=2\bar \theta^{KL\alpha}\psi_{L\alpha}\,,
\cr
\delta\psi_K^\beta&=-i\bar \theta^{IJ\alpha}\epsilon_{IJKL}
(\gamma^\mu)_{\alpha}^{\ \beta} D_\mu\bar C^L\cr & \hskip 1cm 
+\frac{2\pi i}{k}\bar \theta^{IJ\beta}\epsilon_{IJKL} \big(\bar C^LC_P\bar C^P-\bar C^PC_P\bar C^L\big)
+\frac{4\pi i}{k}\bar \theta^{IJ\beta}\epsilon_{IJML}\bar C^MC_K\bar C^L,
\cr
\delta\bar\psi^{K}_\beta&\!=-2i\bar\theta^{KL\alpha}
(\gamma^\mu)_{\alpha\beta} D_\mu C_L\!-\!\frac{4\pi i}{k}\bar\theta^{KL}_\beta(C_L\bar C^M C_M\!-\!C_M\bar C^M C_L)
\!-\!\frac{8\pi i}{k} \bar\theta^{IJ}_\beta C_I\bar C^K C_J,
\nonumber
\end{align}\normalsize
where the parameter $\theta$ was written in terms of $\bar\theta$ using
\begin{equation}
\theta_{IJ}=\frac{1}{2} \epsilon_{IJKL} \bar\theta^{KL}\,.
\end{equation}
Both the supersymmetry parameters are antisymmetric in $I\leftrightarrow J$, and they satisfy the reality condition 
$\bar\theta^{IJ}= (\theta_{IJ})^*$.

\section{Feynman rules}\label{sec:appendixA3}

The \texttt{WiLE} output is written in terms of position-space propagators $\Delta$, which can be computed from those in momentum space \cite{Drukker:2008zx} by means of  the following relation
\begin{equation}
\int \frac{d^{3-2\epsilon} p}{(2\pi)^{3-2\epsilon}} \frac{e^{i p\cdot x}}{(p^{2})^{s}}=\frac{\Gamma\left(\frac{3}{2}-s-\epsilon\right)}{4^{s} \pi^{\frac{3}{2}-\epsilon}\Gamma(s)}\frac{1}{(x^{2})^{\frac{3}{2}-s-\epsilon}}.
\end{equation}
Defining the quantity
\begin{equation}\label{propdelta}
\Delta(x,y)\equiv \frac{\Gamma\left(\frac{1}{2}-\epsilon\right)}{4 
\pi^{\frac{3}{2}-\epsilon}}\frac{1}{((x-y)^{2})^{\frac{1}{2}-\epsilon}}\,,
\end{equation}
in Landau gauge, we have the following propagators
\begin{equation}
\begin{split}\label{propABJM}
\langle (A_{\mu})_{i}^{\ j}(x)  (A_{\nu})_{k}^{\  l} (y)\rangle=&-\left(\frac{2\pi i}{\kappa}\right)\delta_{i}^{l}\delta_{k}^{j}\epsilon_{\mu\nu\rho}\,\partial_{x}^\rho\Delta(x,y)\,,\\
\langle (\hat A_{\mu})_{\hat i}^{\ \hat j}(x)  (\hat A_{\nu})_{\hat k}^{\ \hat l} (y)\rangle=&\left(\frac{2\pi i}{\kappa}\right)\delta_{\hat i}^{\hat l}\delta^{\hat j}_{\hat k}\epsilon_{\mu\nu\rho}\,\partial^{\rho}_{x}\Delta(x,y)\,,\\
\langle (C_{I})_{i}^{\ \hat j}(x)  (\bar C^{J})_{\hat k}^{\  l} (y)\rangle=&\; \delta^{J}_{I}\delta_{i}^{l}\delta_{\hat k}^{\hat j} \,\Delta(x,y)\,,\\
\langle(\psi_{I})_{\hat i}^{\  j}(x)(\bar\psi^{J})_{ k}^{\ \hat l}(y)\rangle=&-i\,\delta^{J}_{I}\delta_{\hat i}^{\hat l}\delta_{ k}^{ j} \,\slashed{\partial}_x\Delta(x,y)\,,\\
\langle (c)_{i}^{\ j}(x)  {(\bar c)}_{k}^{\  l} (y)\rangle=&-\left(\frac{4\pi}{\kappa}\right)\delta_{i}^{l}\delta_{ k}^{ j} \,\Delta(x,y)\,,\\
\langle (\hat{c})_{\hat i}^{\ \hat j}(x) (\bar{\hat{c}})_{\hat k}^{\ \hat l} (y)\rangle=&\left(\frac{4\pi}{\kappa}\right)\delta_{\hat i}^{\hat l}\delta_{\hat k}^{\hat j} \,\Delta(x,y)\,.
\end{split}
\end{equation}

\section{From \texttt{WiLE} output to physical variables}\label{sec:appendixB}

In this section we present the complete list of the symbols appearing in the \texttt{WiLE} output and their physical meaning.
\paragraph{Feynman diagrams}$\\$

\begin{enumerate}
\item Wilson loop contour
\begin{itemize}
\item $\vcenter{\hbox{\includegraphics[trim=1.5cm 5.8cm 0 5.8cm, clip=true, scale=.7]{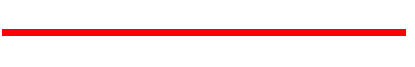}}}\qquad\longrightarrow\quad$ Wilson loop contour line,
\item $\vcenter{\hbox{\includegraphics[trim=1.5cm 0 0 0, clip=true, scale=.7]{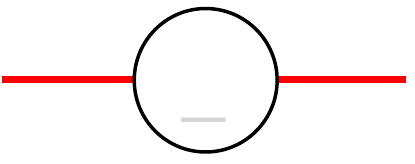}}}\qquad\longrightarrow\quad$ Additional vertex only for graphical purposes when \phantom{x}\qquad\qquad\qquad\qquad\qquad\qquad\,\,the number of fields on the loop is even.
\end{itemize}
\item Fields and vertices positions
\begin{itemize}
\item $\vcenter{\hbox{\includegraphics[trim=1.5cm 0 0 0, clip=true, scale=.7]{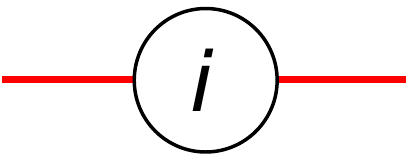}}}\quad\longrightarrow\quad$ Position of the fields on the loop $x_i$,
\item	 $\vcenter{\hbox{\includegraphics[trim=1.5cm 0 0 0, clip=true, scale=.7]{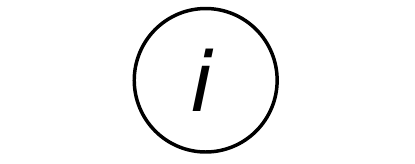}}}\quad\longrightarrow\quad$ Position of the vertices outside the loop $z_i$.
\end{itemize}
\item Propagators
\begin{itemize}
\item $\vcenter{\hbox{\includegraphics[trim=1.55cm 5.8cm 0.24cm 5.8cm, clip=true, scale=.7]{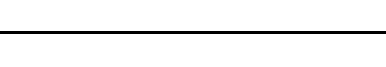}}}\quad\,\longrightarrow\quad$ Fermionic propagator,
\item $\vcenter{\hbox{\includegraphics[trim=1.5cm 5.8cm 0 5.8cm, clip=true, scale=.7]{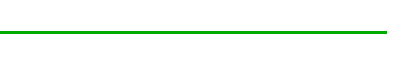}}}\quad\longrightarrow\quad$ Gauge propagator,
\item $\vcenter{\hbox{\includegraphics[trim=1.5cm 5.8cm 0 5.8cm, clip=true, scale=.7]{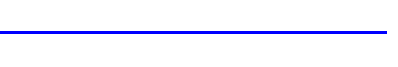}}}\quad\longrightarrow\quad$ Scalar propagator,
\item $\vcenter{\hbox{\includegraphics[trim=1.55cm 5.8cm 0.24cm 5.8cm, clip=true, scale=.7]{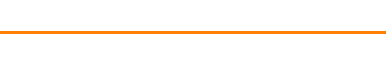}}}\quad\,\longrightarrow\quad$ Ghost propagator.
\end{itemize}
\end{enumerate}

\paragraph{Diagrams integral representation}$\\$

Each entry of the following legend is organized as follows
\begin{equation*}
\text{\texttt{WiLE} output form}\longrightarrow\text{Mathematica full form}\longrightarrow\text{Physical vairable}
\end{equation*}
Notice that the package functions contain many greek characters \texttt{\symbol{92}[}\textit{name}\texttt{]}. They can be typed into the \textit{Mathematica}\textsuperscript{\textregistered} notebook using the shortcut \texttt{Esc+}\textit{name}\texttt{+Esc}. 
\begin{enumerate}
\item The Chern-Simons level $\kappa$ and the gauge group ranks $N$ and $M$
\begin{itemize}
\item $\vcenter{\hbox{\includegraphics{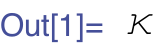}}}\quad\longrightarrow\quad$ \texttt{\symbol{92}[Kappa]}$\,\quad\qquad\longrightarrow\quad\kappa$,
\item $\vcenter{\hbox{\includegraphics{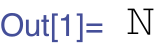}}}\quad\longrightarrow\quad$ \texttt{\symbol{92}[CapitalNu]}$\quad\longrightarrow\quad N$,
\item $\vcenter{\hbox{\includegraphics{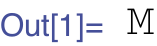}}}\quad\longrightarrow\quad$ \texttt{\symbol{92}[CapitalMu]}$\quad\longrightarrow\quad M$.
\end{itemize}
\item Lorentz\footnote{The same for the other Lorentz indices $\nu_i$ and $\sigma_i$ in the \texttt{WiLE} output.} and R-symmetry indices $\mu_i$ and $J_i$
\begin{itemize}
\item $\vcenter{\hbox{\includegraphics{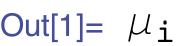}}}\quad\longrightarrow\quad$ \texttt{\symbol{92}[Mu][i]}$\quad\longrightarrow\quad \mu_i$,
\item $\vcenter{\hbox{\includegraphics{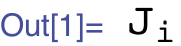}}}\quad\longrightarrow\quad$ \texttt{J[i]}$\qquad\quad\longrightarrow\quad J_i$.
\end{itemize}
\item The position of the fields on the Wilson loop contour $x_i$, the derivative of the position respect to $\tau_i$ $\dot{x}_i^\mu$, its module $|\dot{x}_i|$ and the vertices position $z_i$
\begin{itemize}
\item $\vcenter{\hbox{\includegraphics{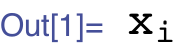}}}\;\;\,\quad\qquad\quad\longrightarrow\quad$ \texttt{x[\symbol{92}[Tau][i]]}$\qquad\;\;\quad\quad\longrightarrow\quad x_{i}$,
\item $\vcenter{\hbox{\includegraphics{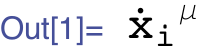}}}\quad\qquad\quad\longrightarrow\quad$ \texttt{xp[\symbol{92}[Tau][i],\symbol{92}[Mu]]}$\quad\longrightarrow\quad \dot{x}_i^\mu$,
\item $\vcenter{\hbox{\includegraphics{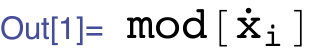}}}\quad\longrightarrow\quad$ \texttt{mod[xp[\symbol{92}[Tau][i]]]}$\;\;\quad\longrightarrow\quad |\dot{x}_i|$,
\item $\vcenter{\hbox{\includegraphics{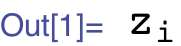}}}\qquad\qquad\;\,\,\longrightarrow\quad$ \texttt{z[i]}$\;\qquad\qquad\qquad\qquad\;\;\longrightarrow\quad z_i$.
\end{itemize}
\item The Dirac gamma matrices $\gamma^\mu$
\begin{itemize}
\item $\vcenter{\hbox{\includegraphics{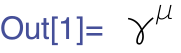}}}\quad\longrightarrow\quad$ \texttt{\symbol{92}[Gamma][\symbol{92}[Mu]]}$\quad\longrightarrow\quad \gamma^\mu$.
\end{itemize}
\item The Levi-Civita tensors with Lorentz indices $\epsilon_{\mu\nu\rho}$ and with R-symmetry indices $\epsilon_{IJKL}$
\begin{itemize}
\item $\vcenter{\hbox{\includegraphics{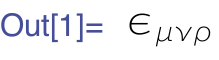}}}\;\!\qquad\longrightarrow\quad$ \texttt{\symbol{92}[Epsilon][\symbol{92}[Mu],\symbol{92}[Nu],\symbol{92}[Rho]]}$\quad\longrightarrow\quad \epsilon_{\mu\nu\rho}$,
\item $\vcenter{\hbox{\includegraphics{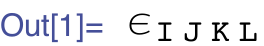}}}\quad\longrightarrow\quad$ \texttt{\symbol{92}[Epsilon]\symbol{92}[Epsilon][I,J,K,L]}$\;\,\quad\longrightarrow\quad \epsilon_{IJKL}$.
\end{itemize}
\item The reduced vector couplings $n_i$ and $\bar{n}_i$
\begin{itemize}
\item $\vcenter{\hbox{\includegraphics{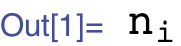}}}\;\,\;\quad\longrightarrow\quad$ \texttt{nJ[\symbol{92}[Tau][i]]}$\;\;\qquad\longrightarrow\quad n_i$,
\item $\vcenter{\hbox{\includegraphics{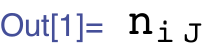}}}\quad\longrightarrow\quad$ \texttt{nJ[\symbol{92}[Tau][i],J]}$\;\,\quad\longrightarrow\quad {n_{i}}_{\, J}$,
\item $\vcenter{\hbox{\includegraphics{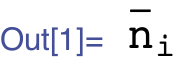}}}\;\,\;\quad\longrightarrow\quad$ \texttt{nJb[\symbol{92}[Tau][i]]}$\qquad\longrightarrow\quad \bar{n}_i$,
\item $\vcenter{\hbox{\includegraphics{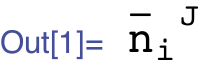}}}\!\;\quad\longrightarrow\quad$ \texttt{nJb[\symbol{92}[Tau][i],J]}$\quad\longrightarrow\quad {\bar{n}_i}^{\;\, J}$.
\end{itemize}
\item The Grassman even fermionic couplings $\eta_i$ and $\bar{\eta}_i$ 
\begin{itemize}
\item $\vcenter{\hbox{\includegraphics{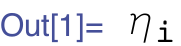}}}\quad\longrightarrow\quad$ \texttt{\symbol{92}[Eta][x[\symbol{92}[Tau][i]]]}$\;\,\quad\longrightarrow\quad \eta_i$,
\item $\vcenter{\hbox{\includegraphics{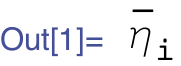}}}\quad\longrightarrow\quad$ \texttt{\symbol{92}[Eta]b[x[\symbol{92}[Tau][i]]]}$\quad\longrightarrow\quad \bar{\eta}_i$.
\end{itemize}
\item The scalar couplings $M_i$ and $\hat{M}_i$
\begin{itemize}
\item $\vcenter{\hbox{\includegraphics{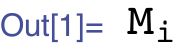}}}\;\qquad\longrightarrow\quad$ \texttt{M[\symbol{92}[Tau][i]]}$\!\!\qquad\qquad\longrightarrow\quad \mathcal{M}_i\quad\text{or}\quad M_i$,
\item $\vcenter{\hbox{\includegraphics{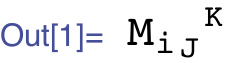}}}\quad\longrightarrow\quad$ \texttt{M[\symbol{92}[Tau][i],J,K]}$\;\;\quad\longrightarrow\quad (\mathcal{M}_i)_J^{\;\;\,K}\quad\text{or}\quad (M_i)_J^{\;\;\,K}$,
\item $\vcenter{\hbox{\includegraphics{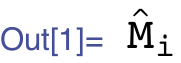}}}\;\qquad\longrightarrow\quad$ \texttt{Mh[\symbol{92}[Tau][i]]}$\,\quad\qquad\longrightarrow\quad \hat{\mathcal{M}}_i\quad\text{or}\quad \hat{M}_i$,
\item $\vcenter{\hbox{\includegraphics{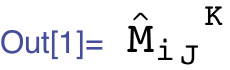}}}\quad\longrightarrow\quad$ \texttt{Mh[\symbol{92}[Tau][i],J,K]}$\quad\longrightarrow\quad (\hat{\mathcal{M}}_i)_J^{\;\;\,K}\quad\text{or}\quad (\hat{M}_i)_J^{\;\;\,K}$.
\end{itemize}
\item The propagator $\Delta$
\begin{itemize}
\item $\vcenter{\hbox{\includegraphics{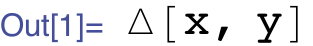}}}\quad\longrightarrow\quad$ \texttt{\symbol{92}[Delta][x,y]}$\quad\longrightarrow\quad \Delta(x,y)$.
\end{itemize}
\item The derivative $\partial_{x}^\mu$
\begin{itemize}
\item $\vcenter{\hbox{\includegraphics{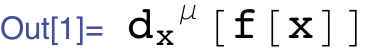}}}\quad\longrightarrow\quad$ \texttt{d[x,\symbol{92}[Mu],f[x]]}$\quad\longrightarrow\quad \partial_{x}^\mu\,f(x)$.
\end{itemize}
\item The product of matrices
\begin{itemize}
\item $\vcenter{\hbox{\includegraphics{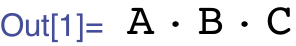}}}\quad\longrightarrow\quad$ \texttt{Centerdot[A,B,C]}$\quad\longrightarrow\quad (ABC)$.
\end{itemize}
\item The trace $\text{tr}$
\begin{itemize}
\item $\vcenter{\hbox{\includegraphics{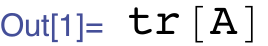}}}\quad\longrightarrow\quad$ \texttt{tr[A]}$\quad\longrightarrow\quad \text{tr}(A)$.
\end{itemize}
\end{enumerate}

\bibliographystyle{nb}

\bibliography{biblio}

\end{document}